\documentclass[12pt]{article}

\input{epsf}
\newcommand{\is}{\! & \!=\! & \! }
\newcommand{\be}{\begin{equation}}
\newcommand{\bea}{\begin{eqnarray}}
\newcommand{\ee}{\end{equation}}
\newcommand{\eea}{\end{eqnarray}}
\newcommand{\ttau}{{\tau}}

\newcommand{\no}{\noindent}
\newcommand{\pa}{\partial }
\newcommand{\omeg}{\mbox{\Large \sc $\omega$}}
\newcommand{\tttau}{\mbox{\large \sc $\tau$}}

\newcommand{\EE}{{\mbox{\tiny $E$}}}

\newcommand{\half}{{\textstyle{1\over 2}}}
\newcommand{\quart}{{\textstyle{1\over 4}}}
\newcommand{\www}{\omega}
\newcommand{\ome}{{\mbox {\small $\omega$}}}
\newcommand{\piome}{{\mbox {\small $\pi \omega$}}}

\newcommand{\non}{\nonumber}

\newcommand{\DD}{{\cal D}}
\newcommand{\rrr}{\rho}

\newcounter{figuur}

\newcommand{\figuur}[4]{
\begin{figure}
[t]
\begin{center}
\leavevmode\hbox{\epsfxsize=#2 \epsffile{#1.eps}}\\[3mm]
\parbox{15.5cm}{\small
\it #3}
\end{center}
\end{figure}
\addtocounter{figuur}{0}
\refstepcounter{figuur}
\label{#4}
}

\catcode`\@=11 \@addtoreset{equation}{section}
\def\theequation{\arabic{section}.\arabic{equation}}
\def\appendix{\renewcommand{\thesection}{\Alph{section}}\setcounter{section}{0}
              \renewcommand{\theequation}
            {\mbox{\Alph{section}.\arabic{equation}}}\setcounter{equation}{0}}

\begin{document}

%%%%%%%%%%%%%%%%%%%%%%%%%%%%%%%%%%%%%%%%%%%%%%%%%%%%%%%%%%%%%%%%%%%%%%%%%%%%
%                      TITLE PAGE                                          %
%%%%%%%%%%%%%%%%%%%%%%%%%%%%%%%%%%%%%%%%%%%%%%%%%%%%%%%%%%%%%%%%%%%%%%%%%%%%
\begin{titlepage}
\begin{center}

{\hbox to\hsize{
\hfill PUPT-2063}}

\bigskip

\vspace{6\baselineskip}

{\Large \bf Schwinger pair creation of Kaluza--Klein particles: 
\vskip .3cm
Pair creation without tunneling$^*$}
\bigskip
\bigskip
\bigskip

{\large Tamar Friedmann\small$^\aleph$ \large and Herman
Verlinde$^\alpha$}\\[1cm]

{ \it Physics Department, Princeton University, Princeton, NJ 08544}\\[5mm]

\vspace*{1.5cm}

{\bf Abstract}\\

\end{center}
\noindent
We study Schwinger pair creation of charged Kaluza-Klein particles from a
static KK electric field.  We find that the gravitational backreaction of the electric
field on the geometry --
which is incorporated via the electric KK Melvin solution -- prevents the
electrostatic potential from overcoming the rest mass of the KK particles,
thus impeding the tunneling mechanism which is often thought of as responsible
for the pair creation. However, we find that pair creation still occurs
with a finite rate formally similar to the classic Schwinger result,
but via an apparently different mechanism, involving a combination of
the Unruh effect and vacuum polarization due to the E-field.

\vfill

%\today
\no $^*$Original title: Schwinger meets Kaluza--Klein 

\no \hskip .2cm Published under present title in Phys.Rev.D71:064018(2005)

\no $^\aleph$ tamarf@feynman.princeton.edu; \hskip .2cm currently at: tamarf@lns.mit.edu

\no $^\alpha$ verlinde@feynman.princeton.edu

\end{titlepage}

\newpage
\section{Introduction and Summary}

\renewcommand{\baselinestretch}{1.08}
\normalsize

The classic study of the rate of creation of electron-positron pairs in
a uniform, constant electric field was done more than 50 years ago in the
seminal paper by Schwinger \cite{Schwinger}. The concepts and
methodology introduced in this work have had a lasting impact
on the formal development of quantum field theory, and by now several
alternative derivations of the effect have been invented (see e.g.
\cite{IZ,AM,Casher:wy,PB}).

Schwinger's predicted rate per unit
time and volume is given by
\cite{Schwinger, PB}
\be \label{schwinger} {\mbox{\footnotesize ${\cal W}$}}(E) \; = \;
{q E} \int \!  {d^2k_i \over {(2\pi)^{2}}} \, \sum
_{n=1}^\infty {1\over n} \exp\Bigl({-{\pi n(m_e^2+k_i ^2)\over |q
E|}}\Bigr)\, \ee for a spin 1/2 particle in $4$ flat space-time
dimensions, with $m_e$ and $q$ the electron mass and charge, $k_i$
the transverse momenta, and $E$ the electric field.\footnote{The
rate (\ref{schwinger}) is still very small for experimentally
accessible electric fields. For the rate to be appreciable, the
field must be very large, around $E_{crit}=10^{16}eV/cm$. A static
field of this magnitude is difficult to obtain in laboratories,
largely because it is several orders of magnitude above the
electric field that can be sustained by an atom, namely $10^8
eV/cm$. See \cite{Bamber:1999zt} for a recent experiment that
has obtained pair creation from oscillating electric fields, which were studied theoretically in \cite{Brezin:xf}, and see \cite{Ringwald:2001ib} for an upcoming experiment studying pair production from the low-frequency, Schwinger limit of such fields.}

In an earlier set of equally classic papers \cite{Kaluza,Klein},
Kaluza and Klein introduced their unified description of general
relativity and electromagnetism, in which charged particles appear
as quanta with non-zero quantized momentum around a compact extra
dimension. It has of course always been clear that the charged
Kaluza-Klein particles do not have the correct properties to
represent electrons; most notably, the mass of the fundamental KK
particles is equal to (or bounded below by) their charge, while
for the electron the ratio $m_e/q$ is about $10^{-21}$\,! So in
comparison with electrons, KK particles are either very heavy or
have (in a large extra dimension scenario) an exceedingly small KK
electric charge. Nonetheless, or rather, because of this fact, it
is an interesting theoretical question whether it is at all
possible, via an idealized Gedanken experiment,
to pair produce KK particles
by means of the Schwinger mechanism. As far as we know,
this question has not been addressed so far in the literature,
and probably for a good reason: it turns out to be a subtle problem! We will
show that unlike the standard Schwinger pair creation effect, pair
production of KK particles cannot be given the simple and rather
intuitive interpretation of a tunneling mechanism.

Imagine setting up our Gedanken experiment as in fig 1, with two
charged plates with a non-zero KK electric field in between.
As seen from eqn (\ref{schwinger}), to turn on the effect by any appreciable
amount will require an enormous KK electric field, and since Kaluza-Klein
theory automatically includes gravity, the backreaction of the E-field on
space-time will need to be taken into account.
The best analog of a constant electric field in this setting is
the electric version of the Kaluza-Klein Melvin background; the magnetic version was studied recently in \cite{Dowker, CG, CC}\footnote{In \cite{Dowker}, pair production of Kaluza-Klein monopoles from the magnetic Kaluza-Klein Melvin solution was studied. See also \cite{Gutperle:2001mb} for a study of other aspects of the magnetic solutions.}, and in \cite{CC} the electric version also appeared. We will study some of the features of the electric KKM background in section 2. For our problem, the relevant properties
of this background are that

\figuur{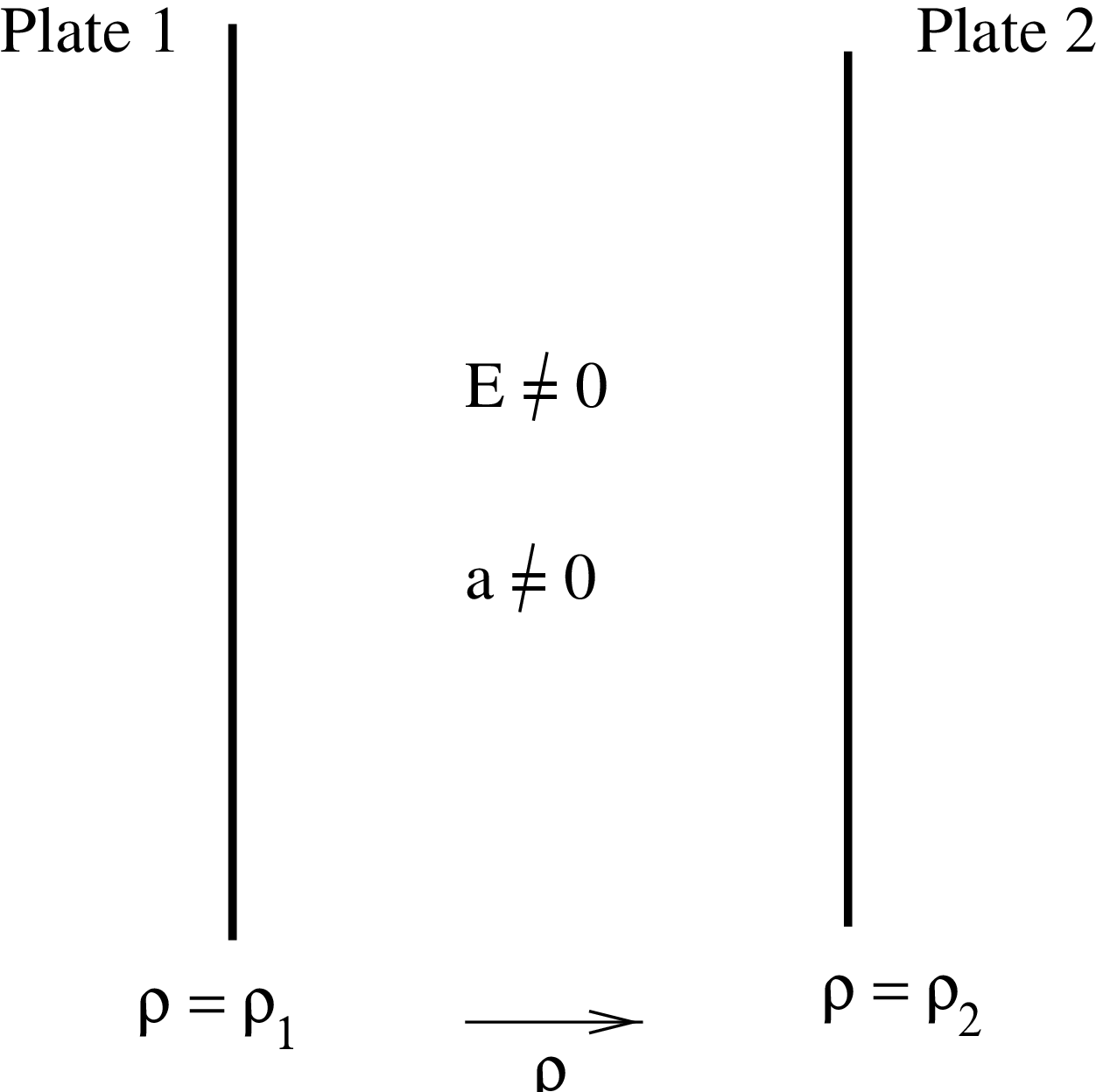}{6cm}{{\bf Fig {\bf \ref{sit0}}:}{\ The Gedanken
experiment we will imagine in this paper, with two charged plates
at $\rho \! = \! \rho_1$ and $\rho \! = \! \rho_2$,  producing a
non-zero E-field in the intermediate region. The backreaction and
the finite mass density of the plates results in a non-zero
gravitational acceleration $a$.} }{sit0}

\noindent
(i) \  the background geometry depends on a longitudinal coordinate, which
we will call $\rho$;

\noindent
(ii) \ a gravitational acceleration $a (\rho)$, directed
along the E-field, is included;

\noindent
(iii) the total gravitational and electrostatic potential energy
remains positive everywhere.

\vskip .2cm \noindent The first two properties are expected
backreaction effects. The last property, however, implies that the
negative electrostatic potential can never be made large enough to
compensate for the positive contribution coming from the rest mass
of the particles. The physical reason for this obstruction is that
before one reaches the critical electro-static potential,
backreaction effects will cause space-time itself to break down:
if one would formally continue the solution beyond this point, the
space-time develops closed time-like curves, which are known to be
unphysical. In this way, gravity puts an upper limit on the
potential difference one can achieve between the two plates in fig
1.

This result may look like an insurmountable obstacle for pair
creation, which is usually \cite{Casher:wy,PB} thought of as a tunneling effect by
which particle pairs can materialize by using their electro-static
energy to overcome their rest mass. The modern instanton method
\cite{AM, Casher:wy} of computing the pair creation rate, for
example, crucially depends on this intuition. However, as
mentioned in point (ii) above, it turns out that the backreaction
necessarily implies that the vacuum state of the KK particles
needs to be defined in the presence of a non-zero gravitational
acceleration. As we will explain in Appendix A, the necessary
presence of this acceleration can be thought of as due to the
non-zero mass of the parallel plates that produce the KK electric
field. Consequently, the Schwinger effect needs to be studied in
conjunction with its direct gravitational analog, the equally
famous Hawking-Unruh effect \cite{Hawking,Unruh,BD}.

It has been recognized for some time that the Hawking-Unruh effect
and Schwinger pair creation are rather closely related (see, for
example, \cite{PB}); both can be understood via a distortion of
the vacuum, which may be parametrized by means of some appropriate
Bogolyubov transformation that relates the standard energy
eigenmodes to the new energy eigenmodes in the non-trivial
background field. Also, like the Schwinger effect, the Hawking--Unruh effect has been thought of as a tunnelling mechanism and was derived as such recently \cite{Parikh:1999mf}; see also \cite{Parikh:2002qh} for a related study of de Sitter radiation.

By combining both the Schwinger and the Unruh 
effects we will obtain the following result for the pair creation
rate of the Kaluza-Klein particles (which we will assume to be
scalar particles) as a function of the electric field $E$ and
gravitational acceleration $a$
%\be \label{uschwinger}
%{\mbox{\footnotesize ${\cal W}$}}(E,a)
%\; = \; { %{\mbox{\footnotesize $\partial_\rho a(\rho)$}
%a \over %\mbox{\footnotesize{
% R}
%\,\sum_{q } \;
% \int {\prod\limits^{{}_{{}^{\! d-2}}}  dk_i
%\over {(2\pi)^d}}
%\; \log\, \Bigl( \, {1 -  e^{-2\pi {\textstyle \omega}(\rho, q, k_i)}}\,
%\over 1 \, - e^{-2\pi {\textstyle \omega}_0(k_i,q)}}\,
%\Bigr)
%\ee

\be \label{uschwinger} {\cal W} (E,a) \, = \, {a^{3/ 2}
\over 2\pi^2   }\, \int \! { {\prod\limits^{{}_{{}^{\! d-2}}}
dk_i \over {(2\pi)^{d-2}}}} \, \sum_{q} \, \left (q^2+\Lambda
k_i^2 \right )^{1/4}
%\sum _{n=1}^\infty \, {1\over n^{3/ 2}}
 \,  \exp\Bigl[- {2\pi}   \, \omeg(a, q, k_i)
\Bigr] \ee where \bea \label{intropot} \omeg(a ,q, k_i) &=& {q^2 +
k_i^2 \over |\, \half \, q E +  a \sqrt{q^2 + \Lambda k_i^2}\,| }\, ,
\qquad \quad  \Lambda = 1 - {E^2 \over 4 a^2}\, ,
%\omeg_0(k_i,q) =  {1\over a} \, \sqrt{q^2 + k_i^2} \\
\eea
%\bea \label{Aintro} A(qR, a, E, k_i)&=&   \, . \eea
%\be
%\omeg(\rho ,q, k_i) = {q^2 + k_i^2 \over |\, \half q E +  a
%\sqrt{q^2 + \Lambda k_i^2}\, |}\, ,
%\qquad \quad  \Lambda = 1 - {E^2 \over 4 a^2}
%%\omeg_0(k_i,q) =  {1\over a} \, \sqrt{q^2 + k_i^2}
%\ee
Here the summation is over the full KK tower of all possible
charges $q = n/R$ with $n$ integer and $R$ the radius of the extra
dimension, and $a$ is the `bare' acceleration, that the particles
would experience with the $E$-field turned off. While $E$ in this
formula is a constant, $a$ in fact depends on the longitudinal
coordinate via $1/a\! = \! \rho\, +$ const. The potential energy
in (\ref{intropot}) is the manifestly positive quantity we
referred to in property (iii) above. A more detailed explanation
of the result (\ref{uschwinger}) will be given in Section
\ref{paircreation}.

%Below, we briefly illustrate the relation between our result and
%the Schwinger and Unruh effects.

Since in our case mass equals charge %(up to a curvature factor),
the result (\ref{uschwinger}) looks like a reasonable
generalization of the classic result (\ref{schwinger}) of
Schwinger and of Unruh \cite{Unruh, PB}.
%\be \label{unruh} {\cal W}(a)={1\over 4\pi}\int \! { {\prod\limits^{{}_{{}^{\! d-2}}}  dk_i
%\over {(2\pi)^{d-2}}}} \,
%-{1\over 4\pi}\log \left (1-e^{-{4\pi w\over a}}\right )=
%\sum _{n,q} {1\over n}\exp \Bigl (-{2\pi n \omega
%\sqrt{q^2+k_i^2}
%\over a}\, \Bigr).
%\ee
In particular, if we turn off the E-field, our
expression (\ref{uschwinger}) reduces to the Boltzmann factor with
Hawking-Unruh temperature $\beta = 2\pi/a$. Moreover, if we would
allow ourselves to drop all terms containing the acceleration $a$,
the result is indeed very similar to the dominant $n=1$ term in
Schwinger's formula (\ref{schwinger}). However, it turns out that
in our case, the gravitational backreaction dictates that the
acceleration $a$ can not be turned off; rather, it is bounded from
below by the electric field via \be a > |E/2|\, . \ee Our formula
(\ref{uschwinger}) indeed breaks down when $a$ gets below this
value. So in particular, there is no continuous weak field limit
in which our result reduces to Schwinger's answer. We will further
discuss the physical interpretation of our result in the
concluding section, where we will make a more complete comparison
with the known rate \cite{Spindel} for Schwinger production in an
accelerating frame.

This paper is organized as follows. In section \ref{EKKM} we
describe some properties of the electric Kaluza-Klein Melvin
space-time.  In sections \ref{particle} and \ref{waves} we study
classical particle mechanics and wave mechanics in this
background. Finally in section \ref{paircreation}, we set out to
calculate the pair creation rate, using (and comparing) several
methods of computation. Section \ref{discussion} contains some
concluding remarks. We discuss our experimental set-up in Appendix
\ref{experimentalapparatus}, and in Appendix \ref{SKK-SR} we
summarize the known result for Schwinger pair production in an
accelerating frame.

\section{The Electric Kaluza-Klein-Melvin Space-Time}\label{EKKM}

We start with describing the classical background of
d+1-dimensional Kaluza-Klein theory, representing a maximally
uniform KK electric field.

\subsection{Definition of the Electric KKM Space-time}

Consider a flat $d+1$ dimensional flat Minkowski space-time, with
the metric \be ds^2 = -dt^2 + dx^2 + dy_i dy^i + dx_{d+1}^2, \ee
with $i=2, \ldots, d-1$. From this we obtain the electric
Kaluza-Klein-Melvin space-time by making the identification \be
\label{map}
\left(\! \begin{array}{c}{t}\\[1mm]{x}\\[.5mm] {y_i}\\[.8mm]x_{d+1}\!
\end{array}\right) \ \ \longrightarrow \ \
%\left(\begin{array}{cc}{\gamma} & {\gamma\beta}\cr{\gamma \beta} & {\gamma}
%\end{array}\right)\,
\left(\! \begin{array}{c}{t'}\\[1mm]{x'} \\[.5mm] {y'_i}\\[.8mm] {x'_{d+1}}
\end{array} \! \right) \; = \;
\left(\! \begin{array}{c}
{\gamma\, (t  - \beta x)}\\[1.2mm] {\gamma\, (x-\beta t)} \\[.5mm] {y_i}
\\[.8mm] {x_{d+1} + 2\pi R}
\end{array}
\! \right), \ee with $\gamma^2(1-\beta^2)=1$. This geometry can be
viewed as a non-trivial Kaluza-Klein background in $d$ dimensions,
in which the standard periodic identification $x_{d+1} \equiv
x_{d+1} + 2\pi R$ of the extra dimension is accompanied by a
Lorentz boost in the $x$-direction. Since the $d+1$ dimensional
space-time is flat everywhere, and the identification map
(\ref{map}) is an isometry, it is evident that the electric Melvin
background solves the equation of motion of the Kaluza-Klein
theory. As we will describe momentarily, from the $d$-dimensional
point of view,  it looks like a non-trivial background with a
constant non-zero electric field $E$ and with, as a result of its
non-zero stress-energy, a curved space-time geometry. Here the
electric field $E$ is related to the boost parameters $\beta$ and
$\gamma$ by \be \label{betaE} \beta  = \tanh (\pi R E)\, , \qquad
\gamma  =  \cosh(\pi R E). \ee

The map (\ref{map}) represents a proper space-like identification,
for which \be -(t'-t)^2 + (x'-x)^2 + (x'_{d+1}-x_{d+1})^2 = (2\pi
R)^2-(2\gamma-2)(x^2-t^2) > 0 \ee provided we restrict to the
region \be \label{region} \rho \, <  {\pi R \over \sinh {\pi ER\over 2}} \,
, \qquad \ \ \rho^2  \equiv \, x^2 - t^2. \ee where we used
(\ref{betaE}). Outside of this regime, the electric Melvin
space-time contains closed time-like curves. We will exclude this
pathological region from our actual physical set-up. \footnote{
There is also a different notion of the electric version of the KK
Melvin space-time, which is obtained by applying an electro-magnetic
duality transformation $F\to e^{2\sqrt{3}\phi} *F$, $\phi \to -\phi$
to the magnetic KK Melvin space-time \cite{Emparan:2001gm}. This background looks
like an electric flux-tube in a $U(1)$ gauge theory with an electric 
coupling constant $e$ that diverges at large transverse distance from 
the flux-tube (due to the fact that the size of the extra dimension
shrinks at large distance). Putting a reasonable physical upper bound on the 
size of $e$ restricts the maximal allowed length of the flux tube, suggesting 
that the obstruction against creating an arbitrarily large 
electro-static potential may be more general than only for the type
of backgrounds studied in this paper. 
}

\subsection{Classical Trajectories}

As a first motivation for the identification of $E$ with the KK
electric field, it is instructive to consider classical
trajectories in this space-time. This is particularly easy, since
in flat $d+1$ Minkowski space, freely moving particles move in
straight lines: \be x = x_0+p_1 s\, , \qquad x^- = t_0+p_0 s \,
,\qquad y_i = k_i s\, , \qquad x_{d+1} = qs\, . \ee Assuming the
particle is massless in $d+1$-dimensions, we have \be \label{msh}
p_0 = \sqrt{p_1^2 + k_i^2 + q^2} \ee which is the mass-shell
relation of a $d$-dimensional particle with mass equal to $q$.
Let's introduce coordinates $\rho$ and $\tau$ via \be
\label{rhodef} x= \rho \, \cosh({\tau - \half E x_{d+1}}) ~,\qquad
\quad t =\rho \, \sinh(\tau - \half Ex_{d+1}). \ee
%\be
%x= \rho \, \cosh({\tau - \half E x_{d+1}})~, \qquad \quad
%t =\rho \, \sinh(\tau - \half Ex_{d+1}).
%\ee
%and
and coordinates $X$ and $T$ by
%\beX^+= X+T= \rho \, e^{\tau} \qquad \quad X^-=X-T = \rho \, e^{-\tau }.\ee
\be \label{newc} X = \rho \, \cosh {\tau} \qquad \quad T = \rho \,
\sinh {\tau }. \ee The identification (\ref{map}) in the new
coordinates becomes \be \label{map4}
\left(\! \begin{array}{c}{T}\\[.5mm]{X}\\[.5mm] {y_i}\\[.8mm]x_{d+1}\!
\end{array}\right) \ \ \longrightarrow \ \
%\left(\begin{array}{cc}{\gamma} & {\gamma\beta}\cr{\gamma \beta} & {\gamma}
%\end{array}\right)\,
\left( \begin{array}{c}{T %+ \pi R E
}\\[.5mm] X \\[.5mm] {y_i}
\\[.8mm] {x_{d+1} + 2\pi R}
\end{array}
\! \right), \ee which is the standard Kaluza-Klein identification.
The trajectory in terms of these is: \bea X\is
%\frac{1}{2 (1+Ex_{d+1})+x^- (1-Ex_{d+1}) \Bigr )
(x_0+p_1 s) \cosh(\half Eqs) + (t_0+ p_0 s  )\sinh(\half Eqs) \non \\[3mm]
T\is
%\frac{1}{2}\Bigl ( x^+ (1+Ex_{d+1})-x^- (1-Ex_{d+1}) \Bigr )
%\frac{1}{2 (1+Ex_{d+1})+x^- (1-Ex_{d+1}) \Bigr )
(t_0+p_0 s) \cosh(\half Eqs) + (x_0+ p_1 s  )\sinh(\half Eqs) \eea
%For small $s$ this reads
%\bea X\is x_0+ ( p_1+Eqt_0  )s+Eqp_0s^2 \non \\T\is
%t_0+ ( p_0+Eqx_0 ) s +Eqp_1s^2
%\eea
Considering a particle at rest at the origin $x_0=0$ and $t_0=0$,
we find
\be \frac{d^2X}{dT^2}=\frac{q E}{p_0} \ee This is the
expected acceleration of a particle with charge and rest-mass $q$.
%In the limit of small $E$, this is
%\be \frac{d^2X}{dT^2}=\frac{2E}{p_0}
%\ee
%\vskip 3cm[The metric in these coordinates is
%\be ds^2=dX^+dX^-+E(X^-dX^+-X^+dX^-)dx_{d+1}+(1-E^2X^+X^-)
%dx_{d+1}^2+dy_mdy^m ~~]\ee

\subsection{Kaluza-Klein Reduction}

Let us now perform the dimensional reduction to $d$ dimensions.
Using the coordinates $\rho$ and $\tau$ defined in (\ref{rhodef})
%\be
%x+t = \rho \, e^{\tau
%+\half \, E\, x_{d+1}} \qquad \quad x-t = \rho \, e^{-\tau -
%\half Ex_{d+1}}.
%\ee
the $d+1$ dimensional metric becomes \be \label{metricE}ds^2=-\rho
^2(d\tau + \half\, E\, dx_{d+1})^2 +
d\rho ^2 +dy_idy^i +dx_{d+1}^2, %\\[3.5mm]
\ee while the identification (\ref{map}) simplifies to a direct
periodicity in $x_{d+1}$ with period $2\pi R$, leaving
$(\rho,\tau,y_i)$ unchanged.
%\be
%\label{map3}
%\left(\! \begin{array}{c}{\tau}\\[.5mm]{\rho}\\[.5mm] {y_i}\\[.8mm]x_{d+1}\!
%\end{array}\right) \ \ \longrightarrow \ \
%\left(\begin{array}{cc}{\gamma} & {\gamma\beta}\cr{\gamma \beta} & {\gamma}
%\end{array}\right)\,
%\left( \begin{array}{c}{\tau %+ 2\pi R\haf E
%}\\[.5mm] \rho \\[.5mm] {y_i}
%\\[.8mm] {x_{d+1} + 2\pi R}
%\end{array}
%\! \right).\ee
We may rewrite the metric (\ref{metricE}) as \be \label{melvind}
ds^2 =  - \, %\Bigl(
\frac{\ \rho ^2}{\Lambda} \, %\Bigr)
d\tau^2 +d\rho ^2 +dy_idy^i+\Lambda \Bigl( dx_{d+1}\! -\frac{E\rho
^2 }{2 \Lambda}\, d\tau \Bigr)^2 \ee with \be \Lambda \, \equiv\,
1 -  {\textstyle{1\over 4}}\, E^2 \rho^2. \ee In this form, we can
readily perform the dimensional reduction.

The $d$ dimensional low energy effective theory is described by
the Einstein-Maxwell theory coupled to the Kaluza-Klein scalar $V$
via \be S = \int \! \sqrt{-g_d}\, \Bigl(\, V^{1/2} R_d + {1\over
4}\, V^{3/2} F_{\mu\nu}F^{\mu\nu}\Bigr). \ee Here the
$d$-dimensional fields are obtained from the $d\!+\!1$ metric via
the decomposition
\newcommand{\Llambda}{V}
\be \label{kk} ds^2_{d+1} = ds^2_{d} + V (dx_{d+1}+ A_\mu dx^\mu)^2
\ee Comparing (\ref{kk}) and (\ref{melvind}) gives the
dimensionally reduced form of the electric Melvin background \bea
\label{emelvin} ds_d^2 \is - \frac{\rho ^2}{\Lambda}d\ttau^2
+d\rho^2 +dy_idy^i \label{g10} \\[2.5mm]
A_{0}\is {E\rho ^2 \over 2 \Lambda} \, , \qquad \quad V \, = \,
\Lambda \, \equiv \, 1- \quart \rho ^2 E^2 ~.\label{A}
% A^{\ttau}=g^{\ttau \ttau}
%A_{\taun}=-E \hskip 1cm A_{\ttau}A^{\ttau}=-E^2 \rho ^2/\Lambda \label{AA}
\eea It describes a curved space-time, together with an electric
field in the $\rho$-direction given by \be E_\rho \, \equiv \,
\sqrt{g^{00}} \, \partial_\rho A_0 \, = \; { E\ \over
\Lambda^{3/2}}. \ee This electric field is equal to $E$ at
$\rho=0$, but diverges at $\rho = 2/E$; this singular behavior is
related to the mentioned fact that outside the region
(\ref{region}), the identification map (\ref{map}) becomes
time-like and produces closed time-like curves. Note, however,
that the location of the divergence in $E_\rho$ slightly differs
from the critical value noted in (\ref{region}), but coincides
with it in the limit of small $E R$.

\figuur{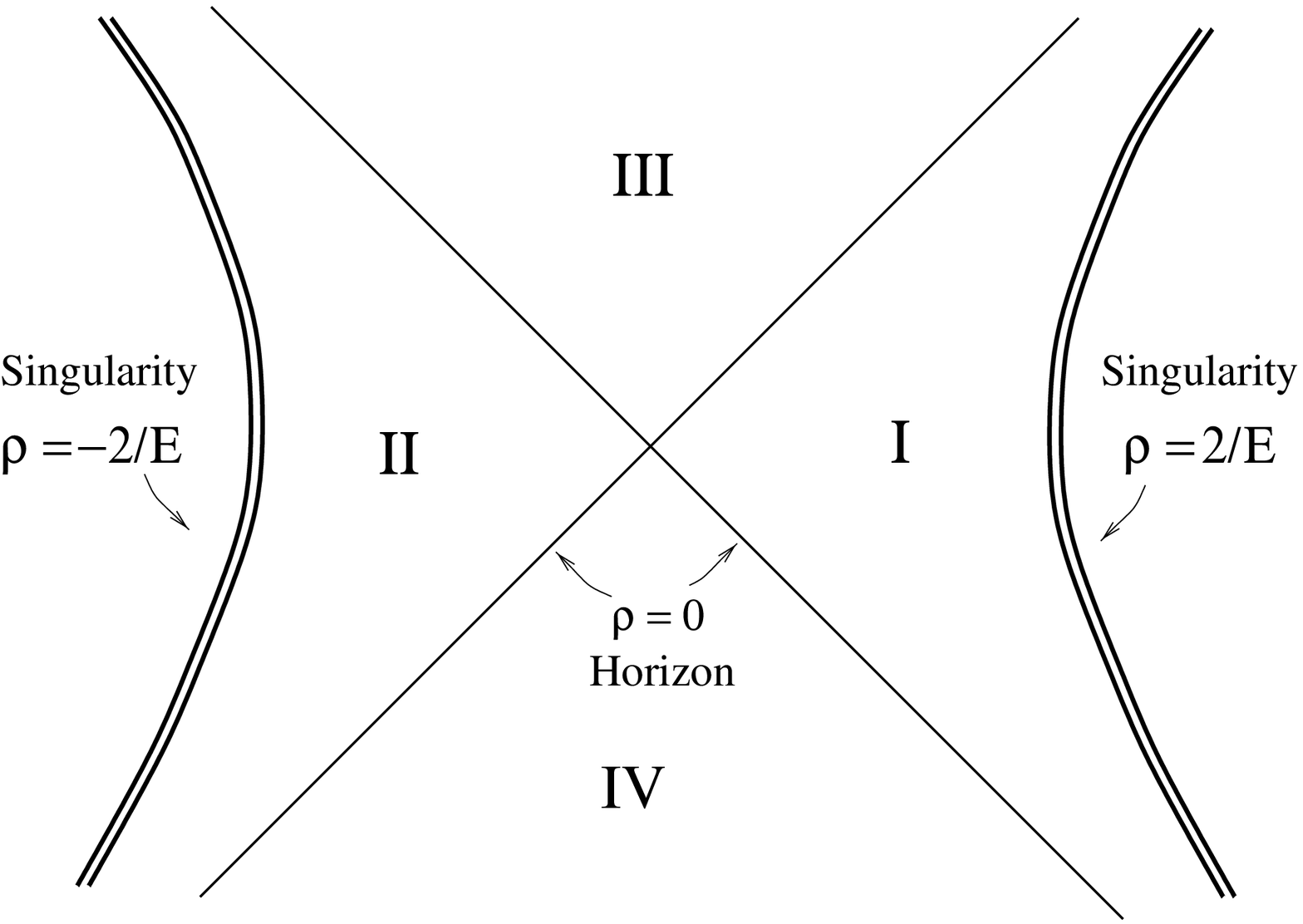}{9cm}{{\bf Fig 2:} The electric KK Melvin space-time
divides up into four regions. Regions I and II are static regions,
while regions III and IV are time-dependent.}{kkm}

The $d$ dimensional metric in (\ref{emelvin}) reduces for $E=0$ to
the standard Rindler space-time metric. For finite $E$ there is a
non-zero gravitational acceleration
\be
a_\rho(\rho) \, \equiv \,
g^{00} \partial_\rho g_{00} \; =\; {1 \over \rho \Lambda}, %1-E^2\rho^2}
\ee which includes the gravitational backreaction due to the
stress-energy contained in the electric field. Notice that
$a(\rho)$ diverges at $\rho=2/E$.

%\figuur{fig2}{7.5cm}{{\bf Fig 3:} The Penrose diagram (taken from
%\cite{CC}) of the full electric KK Melvin space-time.  }

The above static Rindler type coordinate system will be most
useful for the purpose of providing a background with a static KK
electric field. To obtain a more global perspective of the full
electric KK Melvin space-time, we can use the coordinates $X$ and
$T$ defined in eqn (\ref{newc}). In this coordinate system, the
solution looks like \bea ds^2\! \is\! -dT^2 + dX^2 - {E^2\over 4
\Lambda} (XdT -TdX)^2 \, +\, dy_idy^i \, ,
\nonumber \\[3mm]
A_0 \! \is {E X \over 2\Lambda}  \, , \qquad \ \
A_1 \, = -{E T \over 2\Lambda} \, , \\[3mm]
V \! \is \! \Lambda \, , \qquad \ \ \Lambda \equiv  1 - \quart E^2
(X^2 - T^2)\, . \nonumber \eea In this coordinate system we can
distinguish four different regions: \bea {\rm Region \ I}\ :\ \ \
X>|T|\, , \ \ \ & & \ \
{\rm Region \ II\; }: \ \ X<-|T|\, , \nonumber \\[3mm]
{\rm Region \ III}: \ \ T>|X|\, , \ \ \ & & \ \ {\rm Region \ IV}:
\ \ T<-|X| \, .\nonumber \eea
%\be {G}^{(d+1)}=\left ( \begin{array}{cc} {g}^{(d)}g+
%\Llambda A_{\mu}^2 & \Llambda A_{\mu}\\
%\Llambda A_{\mu} & \Llambda \end{array}\right ) \ee
Regions I and II are static regions (that is, they admit a
time-like Killing vector) and are analogous to the left and right
wedges of Rindler space. They are separated by a ``horizon'' (as
seen only by static observers at $\rho=$ const.) at $X^2=T^2$
from two time-dependent regions III and IV. (See fig 2) We will
mostly dealing with the physics of region I. For a discussion of
the physics in region III, see \cite{CC}.

\subsection{Physical Boundary Conditions}

In order to have in mind a physical picture of the part of this
spacetime that we will be studying, we recall the Gedanken experiment
as shown in fig \ref{sit0}, in which two charged plates
produces a {\it static} KKM electric field between them. As
explained in detail in Appendix A, the space-time between the two
plates will correspond to a finite interval within region $I$:
\be \rho_1 < \rho <
\rho_2\, , \qquad \ \ {\rm with} \ \ \ 0<\!\rho_1\!<\! \rho_2\! < |2/E|.
\ee
By concentrating on the physics within this region, our physical set-up
will automatically exclude the unphysical regime with the closed
time-like curves, as well as the horizon at $\rho =0$.
The details of this set-up are given in Appendix \ref{experimentalapparatus}.

\section{Particle Mechanics}\label{particle}

In this section we consider the classical mechanics of charged particles
in the electric KK-Melvin space-time, deriving the expression for the total
gravitational and electrostatic potential energy. This discussion will be
useful later on when
we consider the quantum mechanical pair production.

\subsection{Classical Action}\label{classicalaction}

The classical action for a massless particle in $d+1$ dimensions is
\be
\label{een}
S _{d+1}=\int\! ds \,\Bigl[p_{{}_{M}}\dot{x}^{{}^{M}} + \lambda \,
(G^{{}^{MN}}\!\! p_{{}_M}p_{{}_N})\Bigr],
\ee
where $M,N=0, \ldots ,d$ and $\lambda$ denotes the lagrange multiplier
imposing the zero-mass-shell condition $G^{{}^{MN}}\!\! p_{{}_M}p_{{}_N}=0$.
Upon reduction to $d$ dimensions, using the general Kaluza-Klein
Ansatz (\ref{kk}), for which
\be
{G}^{{}^{MN}}%_{{}_{\scriptsize d+1}}
=
\left ( \begin{array}{cc} g^{\mu\nu}\; & -A^{\nu}\\[2mm] - A^{\mu} &\;
{V}^{{}^{-1}}\!\! +A_{\mu}A^{\mu} \end{array}\right ), \ee where
$\mu, \nu = 0,\ldots , d-1$, the action (\ref{een}) attains the
form (here we drop the $x_{d+1}$-dependence) \be \label{anderhalf}
S_d = \int \! ds\,
\Bigl[p_\mu \, \dot{x}^\mu\, %+\, q \dot{x}_{d+1} \,
+ \, \lambda\, \Bigl(\, g^{\mu\nu}(p_\mu-qA_\mu)(p_\nu- qA_\nu) +
{q^2\over V} \, \Bigr)\Bigr] . \ee Here we identified $q=p_{d+1}$.
The $\lambda$ equation of motion gives \be \label{twee}
g^{\mu\nu}(p_\mu-q A_\mu) (p_\nu-q A_\nu) + {q^2\over V} =0 \ee
This is the constraint equation of motion of a particle with
charge $q$ and a (space-time dependent) mass $m =
{q\over\sqrt{V}}$. For the electric KK Melvin background
(\ref{emelvin}), the constraint (\ref{twee}) takes the form \be
\label{ham1}-\frac{\Lambda}{\rho ^2} \Bigr(p_{\ttau}+ {qE
\rho^2\over 2 \Lambda} \Bigr)^2 + p_\rho^2 +p_i^2 + {q^2\over
\Lambda} =0, \ee or \be \label{ham2} -{p_\tau^2 \over \rho^2}  +
(q-\half E p_\tau)^2 + p_\rho^2 + p_i^2 = 0 . \ee Since the
background is independent of all coordinates except $\rho$, all
momenta are conserved except $p_\rho$. Let us denote these
conserved quantities by \be p_\tau = \omega, \ \qquad p_i = k_i.
\ee The constraint (\ref{ham2}) allows us to solve for $p_\rho$ in
terms of the conserved quantities as \be \label{defmu} p_\rho \,
=\, \pm \sqrt{{(\omega^2 /\rho^2)}\,  - {\mu}^2}, \ \ \qquad \ \
{\mu}^2 \equiv \, k_i^2 + (q-\half E \omega)^2 . \ee Using this
expression for $p_\rho$, we can write the total action of a given
classical trajectory purely in terms of its beginning and
endpoints as \be S_\pm(x_2,x_1) \, = \,  \omega \, \tau_{{21}} +
\, k_i \, y^i_{{21}} \,
%+\, q \, (x_{d+1})_{{21}}%2\pi R w
\, \pm \,
 \int\limits_{\rho_1}^{\rho_2}
\! d\rho\, \sqrt{{(\omega^2 / \rho^2)}\,  - \mu^2}.
\ee
Performing the integral gives
\be
S_\pm(x_2,x_1) = S_\pm(x_2) - S_\pm(x_1)\, ,
\ee
with
%where we have $x_{d+1}_ - x_{d+1}_2 = 2\pi R w$, with $w$ some integer winding number.
\be %\int \sqrt{\frac{\omega^2}{\rho^2}-k^2}\; d\rho
%&=& i\sqrt{u^2-w^2}-iw\tan ^{-1}{\frac{\sqrt{u^2-w^2}}{w}}
%\\ &=& i\sqrt{u^2-w^2}-w\log {\frac {w+i\sqrt{u^2-w^2}}{u}}
\label{classact}
S_\pm(\{ \rho,\tau,y \})=  k_i\, y^i \, + \, \omega  \left ( \tau \pm \, \, %\,  q x_{d+1}\, +
\tttau_0(\rho,k _i,\omega)\, \right ),
\ee
where
\be
\label{tttau}
\tttau_{\! 0}(\rho,k _i,\omega) \; =\;
 {\textstyle\sqrt{1- \Bigl({\mu\rho\over \omega}\Bigr)^2}} \; -\;
\log\!\left[ \frac{\omega}{\mu\rho}\, \Bigl(
1 + \sqrt{\textstyle{1-\Bigl(\frac{\mu\rho}{\omega}\Bigr)^2}}\;\Bigr) \right]\, .
\ee
This result will become useful in the following.

Notice that, for given radial location $\rho$, the classical
trajectory only crosses this location provided the energy $\omega$
satisfies $\omega  \geq  \mu \rho$ with $\mu$ as defined in
(\ref{defmu}). The physical meaning of the quantity $\tttau_{\!
0}$ in (\ref{tttau}) is that it specifies the (time difference
between the) instances $\tau = \pm \tttau_{\! 0}$ at which the
trajectory passes through this radial location. Notice that indeed
$\tttau_{\! 0}=0$ when $\omega = \mu\rho$, indicating that at this
energy, $\rho$ is the turning point of the trajectory.

\subsection{Potential Energy}

We can use the mass-shell constraint (\ref{ham1}) to solve for the
total energy \be \label{hdef} H \equiv p_\tau =   {\rho \over
\Lambda} \, \sqrt{\Lambda(p_\rho^2 + k_i^2) \, + \, {q^2 }} \, -
\, {q E \rho^2 \over 2 \Lambda}. \ee The corresponding Hamilton
equations \be
\partial_\tau \rho = {\partial H \over \partial p_\rho}
\, , \qquad
\partial_\tau p_\rho = - {\partial H \over \partial \rho}\, ,
\ee
determine the classical trajectory $\rho(\tau)$.
An important quantity in the following will be the potential energy
$\omeg(\rho,q,k_i)$, defined via
\bea
\label{rpot}
\omeg(\rho,q,
k_i) \, \equiv \,H(p_\rho =0) %\nonumber \\[3.5mm] \is
\; \, = \; \, { \rho \over \Lambda}
 {\sqrt{q^2 + \Lambda k_i^2}}\; - \; {q E \rho^2 \over 2 \Lambda}.
%= {|q|\, \rho \over 1\, -\,   {q\over |q|}\, E \rho}
\eea
That is, $\omeg(\rho,q,k_i)$ is the energy of particles, with
$p_{d+1}\! =\! q$ and transverse momentum $p_i=k_i$, that have their turning
point at $\rho$.

As expected, the potential energy $\omeg(\rho)$ contains two
contributions: the first term is the gravitational energy due to
the rest mass and momentum of the particle, and the second term
represents the electro-static potential. If $qE$ is positive, this
last term makes the particle effectively lighter than its
gravitational energy. The total energy for any $qE$, however,
never becomes negative. For $qE$ negative, the expression
(\ref{rpot}) is manifestly positive. For $qE$ either positive or negative, it can be
rewritten as \be \label{rpot2} \omeg(\rho, q, k_i) = {q^2 + k_i^2
\over |\,\half q E \, +  \rho^{-1} \sqrt{q^2 + \Lambda k_i^2}\,|
}\, , \ee which is again manifestly positive. We have plotted this
function for $k_i^2=0$ in fig \ref{nplot2}.
\figuur{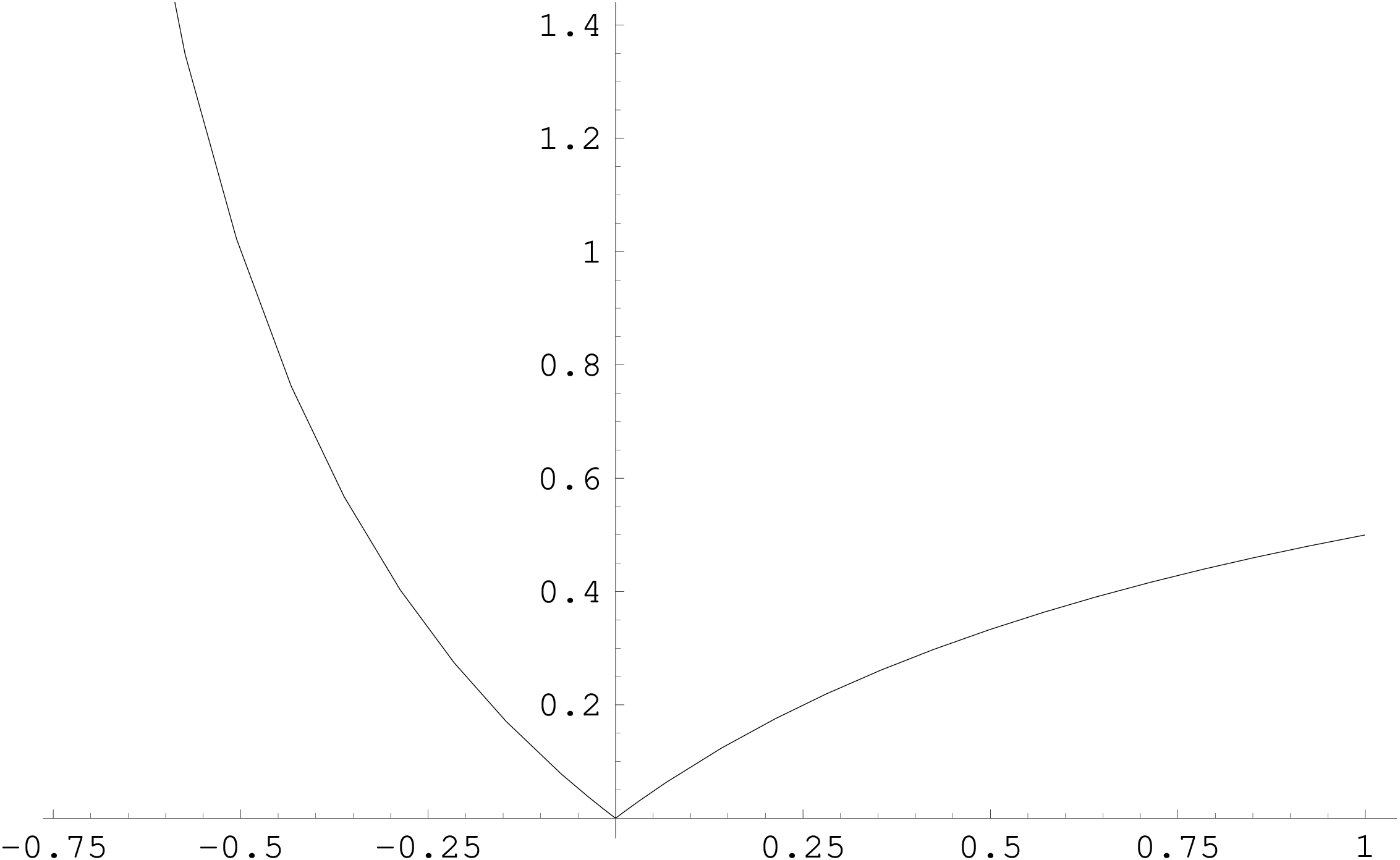}{6cm}{{\bf Fig. \ref{nplot2}} \ {The effective
potential $\omeg(\rho)$ defined in eqn (\ref{rpot}) {\rm (for
$k_i=0$, and multiplied by $E$)} as a function of $x=\half q E
\rho$, with $q=\pm 1$.}}{nplot2}

This behavior of the potential energy $\omeg(\rho , q, k_i)$ should
be contrasted with the classical electrostatic case, where $V(\rho , q) =
m - q E\rho$ with $m$ the rest-mass, in which case the particle
{\it can} get a negative total energy.
When going to single particle wave mechanics, this negative energy
leads to the famous Klein paradox, and upon second quantization,
to the Schwinger pair creation effect. Since in our case the
potential remains positive, there is no Klein paradox and no
immediate reason to expect a vacuum instability. Nonetheless,
as we will see shortly, pair creation will take place.

Finally, we note  that in the concrete set-up of Situation I of our Gedanken apparatus in Appendix \ref{experimentalapparatus}, the
particles are in fact restricted to move within the region
$\rho_1\!<\! \rho\!<\! \rho _2$ between the two plates. To complete
the dynamical rules of the model, we need to specify what happens
when the particle reaches the plates; we will simply assume reflecting
boundary conditions.

\section{Wave Mechanics}\label{waves}

In this section we write the solutions to the wave equations in the electric KK-Melvin
background, and illustrate the semi-classical correspondence with the
classical mehanics.

\subsection{Wave Equations}\label{waveeqns}
The $d+1$-dimensional wave equation in the background (\ref{metricE}) is
\be \label{e1} \frac{1}{\sqrt{-G}}\, \partial_M \! \left ( \sqrt{-G}G^{{}^{MN}}
\! \partial_N \Phi \right ) =
%\left (\pa ^2 _{y_i}+\frac{1}{\rho}\partial _\rho (\rho \partial _\rho ) -\frac{1}{\rho ^2}\partial ^2 _\taun
%+\partial ^2_{d+1} \right )\Phi = 0
\left [ \frac{1}{\rho}\, \partial _\rho
(\rho \, \partial _\rho )-\frac{1}{\rho^2}\, \pa _{\ttau}^2 +\, \pa_i^2\, +\,
\left ( \pa _{d+1}+\half E\pa _{\ttau}\right ) ^2 \right ] \Phi = 0,
\ee
subject to the perioding boundary condition in the $x_{d+1}$ direction
with period $2\pi R$. For a given eigenmode with $q\equiv p_{d+1} =
{n\over R}$, we can reduce the wave equation to
$d$ dimensions, where it can be written in the form
\be \label{gce}\left ( \frac{\sqrt{\Lambda}}{\rho}\,
\partial_\rho
\Bigl( \frac{\rho}{\sqrt{\Lambda}} \,
\partial _\rho \Bigr)\, -\, \frac{\Lambda}{\rho ^2}\Bigl(
\pa _{\ttau} + {i q E\rho^2 \over 2 \Lambda}\Bigr)^2 \,
%e^2 E^2\rho ^2 /\Lambda
\, + \, \pa_i^2 \, -
\, {q^2 \over \Lambda}
\right )\Phi= 0\, .
\ee
Here we recognize the conventional wave equation
\be  \frac{1}{\sqrt{-g}}D _\mu \left ( \sqrt{-g}g^{\mu \nu}D _\nu
\Phi \right )
- M^2 \Phi =0
\ee
of a $d$ dimensional charged particle with charge $q$
in the background (\ref{emelvin}) and with a position dependent
mass equal to $M^2 ={q ^2 / \Lambda}$.
Note the direct correspondence af the above wave equations with the
classical equations (\ref{ham1}) and (\ref{ham2}).
%via the usual identifications:
%\[ i\pa _{\ttau}= p_{\ttau} \hskip 1cm  i\pa _\rho =p_\rho \hskip 1cm
%i\pa _{d+1} =q \hskip 1cm i\pa _{i} = p_i \]
They need to be solved subject to the boundary
conditions imposed by our physical set-up. In case of Situation I,
see fig. \ref{sit1} in Appendix \ref{experimentalapparatus}, we will choose to impose Dirichlet boundary conditions at
the two plates
\be \label{bc}
\Phi~|_{\rho=\rho_1} = \Phi~|_{\rho=\rho_2} = 0\, .
\ee

\subsection{Mode Solutions}

The $d+1$-dimensional wave equation is solved by \be \label{mode}
\Phi_{qk\omega} = e^{i x_{d+1}(q-{1\over 2} E\omega )
%} \Phi_{qk\omega}(x^\mu)
%\eewith\be\Phi_{qk\omega}(x^\mu) =  e^{
+ ik_i y^i+i\omega \ttau} K(\omega, \mu\rho )\, ,
\ee
%\be\mu^2 =  (q + \half \omega E)^2+k_i^2 \ , \qquad \quad
with  $\mu$ as defined in (\ref{defmu}), and
where $K(\omega,\mu\rho)$ solves the differential equation
\be
\label{keq}
\Bigl((\rho\partial_\rho)^2 + \omega^2 + \mu^2 \rho^2\Bigr) K(\omega,\mu\rho) = 0.
\ee
The solution $K$ has the integral representation
%\be J_{i\omega}(i\mu\rho )=\int\limits_0 ^\infty \! \frac{ds}{s}\; s^{i\omega}\,
%e^{\frac {i\mu\rho}{2}(-s+{1/ s})} \ee
%With thechange of variables $s=e^\sigma$ this becomes
\be
\label{kdef}
K(\omega, \mu\rho )=
\, \int\limits_{-\infty}^{\infty}\!\! d\sigma \;
e^{i\omega\, \sigma - {i\mu\rho} \sinh\sigma} \, , %= \int _{-\infty}^\infty d\tau e^{if(\tau)}
\ee
and can be expressed in terms of standard Bessel and Hankel functions
%\cite{GR}
\cite{GR, 'tHooft:1996tq}. The functions $K(\omega , \mu \rho)$ are %all real.They are
defined for arbirary real $\omega$. However, upon imposing the
boundary conditions that $K(\omega,\mu\rho_i) =0$ at the location
of the two plates, we are left with only a discrete set of allowed
frequencies $\omega_\ell$. Since the corresponding mode functions
(\ref{kdef}) form a complete basis of solutions to (\ref{keq}),
they satisfy an orthogonality relation of the form \be \label{inn}
\int\limits_{\rho_1}^{\rho_2}\! d\rho \, \rho \, K^*(\omega_{{}_
{\! \ell}},\mu_{{}_{\! \ell}}\rho) K(\omega_{{}_{\! j}},\mu_{{}_{\! j}}\, \rho) =
f(\omega_\ell) \delta_{\ell,j} \, , \ee
where $f(\omega)$ some
given function that depends on $\rho_1$ and $\rho_2$.

For large $\omega$ and $\mu\rho$, we can approximate the integral
in (\ref{kdef}) using the stationary phase approximation. The
stationary phase condition $\omega = {\mu\rho}\cosh\sigma$ has two solutions
\be
\sigma_\pm \, = \; \pm \log\!\left[ \frac{\omega}{\mu\rho}\, \Bigl(
1 + \sqrt{\textstyle{1-\Bigl(\frac{\mu\rho}{\omega}\Bigr)^2}}\;\Bigr) \right]
\ee
provided $|\omega |\!>\!\mu\rho$, leading to
\be
\label{semc}
\qquad \ \ K(\omega, \mu\rho )
\; \simeq
{\sqrt{2 \pi} \, \cos\Bigl( \omega {\tttau}_{ 0}(\rho)+{\pi \over 4}\Bigr)\over
\sqrt{w}\, \, \, {}^{{}^{\mbox{\scriptsize 4}}}\!\!\!\!\sqrt{
1- \Bigl({\mu\rho\over \omega}\Bigr)^2}}\, , \qquad \ \ {\mbox{\small $
|\omega |> \mu\rho$}} \, ,
\ee
with $\tttau_0$ as given in eqn (\ref{tttau}).  This formula is accurate
for energies $\omega$ larger than the potential energy $\omeg(\rho)$.
For smaller energies there is
no saddle-point and the function $K(\omega,\mu\rho)$ is exponentially
small
\be
\label{exup}
\qquad K(\omega,\mu\rho) \, \simeq \, \sqrt{\pi \over \mu\rho} \;\, e^{-
{\mbox {\small $\mu\rho$}}}
\, , \qquad \qquad \quad {\mbox{\small $
|\omega | <\!\!< { \mu \rho}$}}
\ee
reflecting the fact that the corresponding classical trajectory has
its turning point before reaching $\rho$.

Notice that, upon inserting (\ref{semc}), the full mode function
$\Phi_{qk\omega}$ in (\ref{mode}) can be written as a sum
of two semi-classical contributions
\be
\Phi_{qk\omega}(x)\, \sim \, \sum_\pm %\,c_{{}_{\pm}}
\, e^{ i x_{d+1}(q+{1\over 2} E\omega )+ ik_i y^i + i \omega \,
(\tau \pm {\textstyle\tau}_{\! 0}(\rho,k,\omega))} \; \sim \; \sum_\pm %c_{{}_{\pm}}
e^{i S_\pm(y,\tau,\rho)},
\ee
corresponding to the left- and right-moving part of the trajectory,
respectively.

\vskip 1.3cm

\section{Pair Creation}\label{paircreation}

In this section we will compute the pair creation rate of the Kaluza-Klein
particles, following three different (though related) methods. We will start
with the simplest method, by looking for Euclidean ``bounce'' solutions.
We then proceed with a more refined method of computation,
more along the lines of Schwinger's original calculation, producing the
non-trivial result quoted in the introductory section. Finally, we show
that the obtained result can naturally be interpreted
by considering the Hawking-Unruh effect, and we use the method of Bogolyubov
transformations to compute the expectation value of the
charge current.

\subsection{Classical Euclidean Trajectories}

Assuming that, in spite of the fact that the effective potential (\ref{rpot})
seems to suggest otherwise, the nucleation of the
charged particle pairs can be viewed as the result of a quantum mechanical
tunneling process, we compute the rate by considering
the corresponding Euclidean classical trajectory. The analytic continuation
of the electric KK Melvin space-time to Euclidean space is
\bea
\label{melvine}
ds_\EE^2 \is %\Bigl(
\frac{\ \rho ^2}{\Lambda _\EE} \, %\Bigr)
d\theta^2 +d\rho ^2 +dy_idy^i+\Lambda _\EE
\Bigl( dx_{d+1}\! - \frac{E\rho ^2 }{2 \Lambda _\EE}\, d\theta \Bigr)^2 ,\\[3mm]
A_\theta \is {E\rho^2\over 2\Lambda_\EE}, \qquad \quad V = \Lambda_\EE ~,
\nonumber
\eea
with $\theta$ a periodic variable with period $2\pi$, and
\be
\label{zto} \Lambda_\EE \, \equiv\, 1 + E^2 \rho^2/4 \, .
%\qquad \epsilon=\pm 1.
\ee
This Euclidean geometry is obtained from the Lorentzian electric
KK Melvin solution via the replacement
\be
\label{wick1}
E \to i E\, , \qquad \ \   t \to - i\theta,
\ee
and coincides with the space-like section of the magnetic KK
Melvin space-time. Unlike the Lorentzian version, this Euclidean
space-time extends over the whole range of positive $\rho$ values
and ends smoothly at $\rho=0$, by virtue of the periodicity in
$\theta$. This is standard for Euclidean cousins of space-times
with event horizons, and a first indication that quantum field
theory in the space-time naturally involves physics at a specific
finite temperature.

The Euclidean action of a point-particle, with charge (momentum in
the $d\!+\!1$-direction) equal to $p_{d+1} = q$ and mass
$M=|q|/\sqrt{\Lambda _\EE}$ moving in this background reads \be
\label{seucl} S_\EE \, = \, \int \! ds \, {\cal L}_E\; , \qquad
{\cal L}_E = \, {|q|\over \sqrt{\Lambda_\EE}} \, \sqrt{{\rho^2
\dot{\theta}^2 \over \Lambda_\EE} \, + \dot\rho^2 + \dot{y}_i^2}
\; \; - \; {q E \rho^2 \over 2 \Lambda_\EE} \, \dot \theta %\right)
. \ee As a first step, let us look for closed circular classical
trajectories at constant $\rho$ and $y_i$. The above
point-particle action then reduces to \be \label{seff}
S_\EE(\rho)\, = \, {\pi \over \Lambda_\EE} \, (2 |\, q|\, \rho \,
- q E \rho^2). \ee The first term is the energy of a static
particle times the length of the orbit, and the second term is the
interaction with the background field times the area of the loop.
Looking for an extremum yields
%\be
%\partial_\rho S_\EE(\rho)\, = \, {2\pi\over \Lambda_\EE^2} \, \Bigl( \, |q|
%(1- E^2\rho^2/4) +  qE\rho \Bigr) \; = \; 0\, ,
%\ee
%which has
one real and positive solution \be \label{perhaps} \qquad |E|\rho
= 2 \sqrt{2} - 2 \cdot {\rm sign}(qE)\, \qquad \ee with total
action \be \label{euclid} S_\EE = {2\pi |q| \over |E|} \,
(\sqrt{2} - {\rm sign}(qE)). \ee The existence of these solutions
with finite Euclidean action is a first encouraging sign that pair
creation may take place after all. The answer (\ref{seff}) for the
Euclidean action also looks like a rather direct generalization of
the standard semiclassical action for the Schwinger effect, and
%with $M^2=q^2/\Lambda _\EE$ and $E_\rho =
%E/\Lambda _\EE ^{3/2}$, it can be rewritten as \be S_\EE (\rho) =
%2\pi \rho \left ( {M^2 \over |q|}\right ) + qE_\rho \sqrt{\Lambda
%_\EE} ( \pi \rho ^2  ) \, , \ee which generalizes the action
%$I=2\pi RM-gB\pi R^2$ of \cite{AM} 
it is therefore tempting to conclude at this point that the total
pair creation rate is proportional to \be e^{-S_\EE} \; = \;
e^{-{2\pi |q| \, (\sqrt{2}-{\rm sign}(qE))/ |E|}}, \ee which looks
only like a numerical modification of the classic result
(\ref{schwinger}). This conclusion is somewhat premature, however,
since in particular the pair creation rate should depend on
$\rho$. We would like to determine this $\rho$-dependence.

For this, we take a second step and consider closed
Euclidean trajectories that are not necessarily circular.
As in section \ref{particle}, we now go to
a Hamiltonian formulation. To transform the formulas in
section \ref{particle}  to the Euclidean
set-up, we need to make, in addition to (\ref{wick1}), the following
replacements 
\be \label{wick2} s \to - is\, , \qquad \ \ p_\rho \to i p_\rho,
\qquad \ \ p_i \to i p_i\, , \qquad \ \ q \to iq \, . \ee In this
way we obtain from (\ref{hdef}) a Euclidean Hamiltonian \be
\label{he} H_\EE \equiv p_\theta =   \, {\rho \over \Lambda _\EE}
\, \sqrt{|q|^2 - \Lambda _\EE (p_\rho^2 + k_i^2)}\, - {q E\rho^2
\over 2 \Lambda _\EE} \,, \ee that generates the motion of
particle as a function of the Euclidean time $\theta$, and a
corresponding potential energy \be \label{epot} \omeg_{\!
\EE}(\rho,k_i,q) \equiv -H_\EE (p_\rho =0) = \, - {\rho \over
\Lambda_\EE} \, \sqrt{|q|^2 - \Lambda_\EE k_i^2 }\, + \,{q E\rho^2
\over 2 \Lambda_\EE} \, . \ee In addition to a change in sign,
which is the standard way in which a potential changes when going
to Euclidean space, this Euclidean potential differs from
(\ref{rpot}) via the replacement $\Lambda \to \Lambda_\EE$. We
have drawn $\omeg_{\! \EE}$ for $k_i=0$ in fig \ref{pot5}. Note
that $\omeg_{\! \EE}$ for $k_i=0$ is proportional to the reduced
effective action (\ref{seff}) for circular trajectories, and the
critical radii (\ref{perhaps}) reside at the two minima in fig
\ref{pot5}.

\figuur{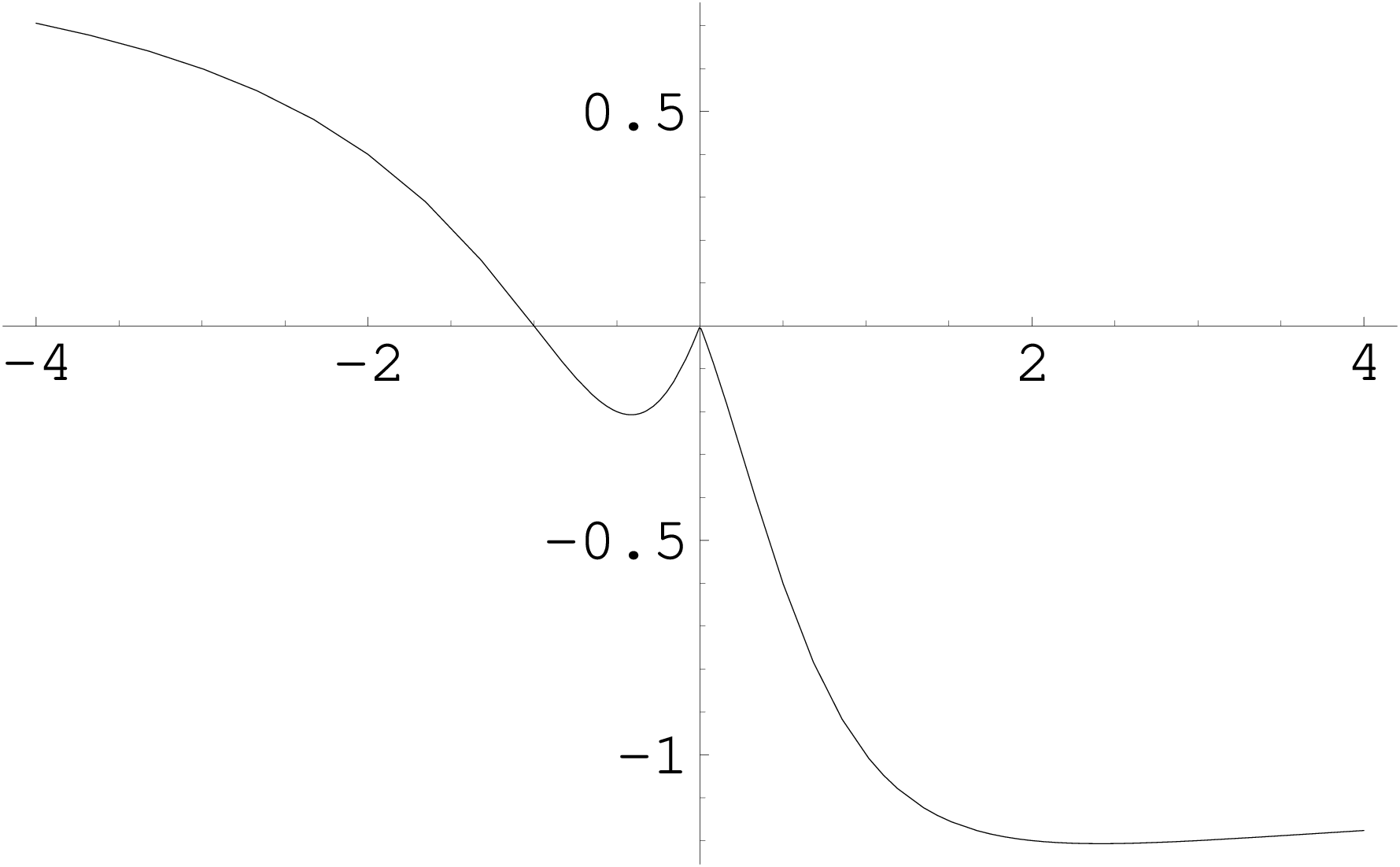}{6.3cm}{{\bf Fig. {\bf \ref{pot5}} } \
{The Euclidean effective potential $\omeg_\EE(\rho)$ defined in eqn
(\ref{epot}) {\rm (for $k_i=0$, and multiplied by $E$)} as a function of
$x=\half q E \rho$, with $q=\pm 1$.
}}{pot5}

Our goal is to obtain semi-classical estimate for the pair creation
rate at some given $\rho$.  How should we use this Euclidean potential
for this purpose?  As seen from fig \ref{pot5}, there is a range of Euclidean
energies $H_\EE$ around the two minima (\ref{perhaps}) for which there
exist stable, compact orbits. These orbits have a maximal and minimal
radius, $\rho_+$ and $\rho_-$, at which $H_\EE = \omeg_\EE(\rho_\pm)$.
The idea now is to associate to a given $\rho$ the
corresponding Euclidean trajectory for which $\rho$ equals one of
these extrema $\rho_\pm$, and then use the total action $S_\EE(\rho)$
for this trajectory to get a semi-classical estimate of the pair
creation rate via
\be
\label{semicl}
{\mbox {\small ${\cal W}$}}(\rho) \simeq e^{-S_\EE(\rho)}\, .
\ee
Here it is understood that in $S_\EE(\rho)$ we undo the rotation $E\to iE$, so
that $\Lambda_\EE \to \Lambda$.  Equation (\ref{semicl}) is then a clear
and unambiguous formula, provided the classical orbit is closed.

In
general, however, the orbits need not be closed: the period of
oscillation does not need to be $2\pi$ or even a fraction or multiple
thereof. How should we {\it define} the total classical action, to be
used in (\ref{semicl}) for such a trajectory?

Our proposal, that perhaps may look {\it ad hoc} at this point but
will be confirmed and justified in the subsequent subsections, is
to take for $S_\EE$ the total action averaged over one full
rotation period of $2\pi$. Concretely, suppose that the compact
trajectory has an ``oscillation period'' $\theta_0$, in which it
goes through a full oscillation starting and returning to its
maximal radial position $\rho\!=\! \rho_+$.  We then define
$S_\EE(\rho)$ as: \be S_\EE(\rho) \, \equiv \, \lim_{\theta\to
\infty} \, {2\pi\over \theta} \, \int\limits_0^{\ \theta} \!
d\theta \, {\cal L}_\EE(\theta,\rho)\, \, =\, {2\pi\over \theta_0}
\, \int\limits_0^{\ \theta_0} \! d\theta \, {\cal
L}_\EE(\theta,\rho). \ee  With this definition, and using the
results in Section \ref{classicalaction}, we can now easily
evaluate $S_\EE(\rho)$. From (the Euclidean analog of) eqn
(\ref{classact}), while noting that $\tttau_0(\rho)=0$ since
$\rho$ is the turn-around point, we obtain
%\check{I still don'tsee how this follows}
\be \label{som} S_\EE(\rho) = 2\pi
\omeg(\rho,q,k_i), \ee with $\omeg$ as given in (\ref{rpot}). Here
we made the replacement $E\to iE$, as prescribed.

The result (\ref{som}) together with (\ref{semicl}) gives our proposed
semi-classical estimate of the pair creation rate as a function of
$\rho$. Clearly, the derivation as presented thus far needs some
independent justification. It also leaves several open questions. In
particular, it is not clear how we should interpret the Euclidean
``bounce'' solutions, given the fact that the real effective potential
(\ref{rpot}) doesn't seem to lead to any tunneling. A better
understanding of the physics that leads to the pair creation seems
needed. In the next two subsections we will present two slightly more
refined derivations of the rate, which will help answer some of these
questions.

\bigskip

\subsection{Sum over Euclidean Trajectories}

We will now evaluate the pair creation rate, per unit time and
volume, by means of the path-integral. Since we expect that this
rate will be a function of longitudinal position $\rho$, we would
like to express the final result as an integral over $\rho$. We
start from the sum over all Euclidean trajectories \be
\label{psum} {\cal W} \; = \; \int \DD p\, \DD x \,
\exp\Bigl({-{1\over \hbar} S[p,x]}\Bigr) \ee defined on flat
d+1-dimensional space with metric and periodicity condition \be
ds_E^2 = dx^* dx + dy_idy^i + dx_{d+1}^2, \ee \be (x, x^*, y_i, \,
x_{d+1}) \, \equiv\, ({e^{i\pi ER} x},\, e^{-i\pi ER} x^*,\, y_i,
\, {x_{d+1} + 2\pi R})\, . \label{map2} \ee In the end we intend
to rotate back to Lorentzian signature, replacing $E \rightarrow
iE$.

We can read the expression (\ref{psum}) as a trace over the
quantum mechanical Hilbert space of the single particle described
by the action (\ref{een}) or (\ref{anderhalf}). The idea of the
computation is to write this as a sum over winding sectors around
the 11-th direction. For each winding number $w$, the closed path
is such that the end-points are related via a rotation in the
$(x,x^*)$-plane over an angle $w\pi E R$. Using this insight, we
can write (\ref{psum}) as \be \label{wsum} \int\! \! d^d x \;
{\cal W}(x) \; = \; R \int\limits_0^\infty \! {d
T\over T }\,\sqrt{{2\pi \over  T}}\,
 \sum_w\, e^{-{\textstyle{1\over 2T}}(2\pi R w)^2}\,
{\rm Tr} \Bigl[\, e^{\pi i w E R J } e^{\textstyle -{T\over 2}
(p^*\! p + p_i^2) }\Bigr] \ee where $T$ denotes the Schwinger
proper time variable, and where $J$ denotes the rotation generator
in the $(x,x^*)$ plane. The exponent in front of the the trace is
the d+1-dimensional part of the classical action of the trajectory
with winding number $w$. To compute the trace, we write it as an
integral over mixed position and momentum eigen states \be {\rm
Tr}\, A \, = \, \int \!\! d^2 x \int \!\!
{\prod\limits^{{}_{{}^{\! d-2}}}  dk_i \over (2\pi)^{d-\!2}}\;\;
 \langle\, x, k_i\, | \, A \, |\, x, k_i\, \rangle\,
\ee

%\be
%\langle x| \; e^{\pi i w E R J} e^{-{T\over 2}
%p^+\! p^-\,}\; |x\rangle
%\; = % \qquad \qquad \qquad \qquad \qquad \qquad \qquad \qquad \qquad
%\nonumber \\[3mm]\qquad \qquad \qquad \qquad
%\textstyle
%{1\over \pi T}\,
%\;
%\exp\Bigl(
%e^{-{\mbox{{ ${\ x^+ x^- \over 2 T}$}
%\small $\!\! (e^{-i\pi  E R w}\!-1)(e^{i\pi  E R w}\!-1)$}}}
%\Bigr)
%\ee

Next we evaluate \be \langle x| \; e^{\pi i w E R J} e^{-{T\over
2} p^*\! p\,}\; |x\rangle
\; = % \qquad \qquad \qquad \qquad \qquad \qquad \qquad \qquad \qquad
%\nonumber \\[3mm]\qquad \qquad \qquad \qquad
%\textstyle
{1\over \pi T}\, \;
%\exp\Bigl(
e^{-{\mbox{{ ${\ x x^* \over 2 T}$} \small $\!\! (e^{\pi i E R
w}\!-1)(e^{-\pi i E R w}\!-1)$}}}
%\Bigr)
\ee where we used the standard formula for the heat kernel in two
dimensions. Inserting this into (\ref{wsum}), we can write the
production rate as an integral over $\rho$ of \be \label{form}
{\cal W}(\rho)  \; = \; %2\pi \int\limits_0^\infty \! d\rho \;
R \int\limits_0^\infty \! {d T\over T }\,\sqrt{{2\pi \over T}}\,
\int \!\! {\prod\limits^{{}_{{}^{}_{\!d-2}}}  dk_i
\over (2\pi)^{{d-\!2}}}
 \; {e^{{-{T\over 2}}\, k_i^2}\over  \pi  T } \;
%\left(
\sum_w \; e^{-{\textstyle{1\over 2T}}{\bigl(%\textstyle{
\mbox{\small {$(2\pi R w)^2 \, + \, 4 \rho^2 {\sin}^2(\pi E Rw/2)$}}
\bigr)}}
%\right)
\ee
which we will interpret as the pair production rate at the
location $\rho$.

Equation (\ref{form}) is an exact evaluation of the Euclidean
functional determinant. To put it in a more useful form, we will
assume that we are in the regime $\rho^2 \! >\!\!>\! T$ (an
assumption that we will be able to justify momentarily), so that
we can simplify the expression by means of the Villain
approximation \be \label{villain} \sum_w e^{-{\textstyle{1\over
2T}} {\bigl(\mbox{\small {$(2\pi R w)^2 + 4 \rho^2 {\sin}^2(\pi E
Rw/2)$}} \bigr)}} \simeq  \sum_{w,n} e^{-{\textstyle{1\over
2T}} \bigl(\mbox{\small {$(2\pi R w)^2 + \rho^2 (\pi E Rw - 2 \pi
n)^2$}}\bigr)}. \ee This replacement essentially amounts to a
semi-classical approximation. The right-hand side can be
re-expressed via the Poisson resummation formula (note here that
the left-hand side below is just a trivial rewriting of the
right-hand side above) \be \label{poisresum} \sum_{w} \! \mbox{\
$e$}^{-\mbox{{\large $ {1\over 2T}$}} \mbox{\small $\Bigl(\Lambda
_\EE \Bigl(2\pi R w\! -\!\! {\mbox{\large ${\pi E \rho^2 n \over
\Lambda _\EE }$}}\Bigr)^2 + {\mbox{\large ${4\pi^2 \rho^2 n^2\over
\Lambda _\EE}$}}
\Bigr)$}} \\[3.5mm]% \right)
= \sqrt{{\mbox{\raisebox{-1pt}{\small $T$}}}
\over
{\mbox{\raisebox{-1pt}{\small $2\pi \Lambda_\EE R^2$}}}} \;
\sum_{m} \mbox{$e$}^{-\mbox{\large ${1\over \Lambda_\EE}\bigl({T\over 2}
{m^2 \over R^2}\, {\mbox{\small +}}\, {2 \pi^2 \rho^2 n^2 \over T}
\, \mbox{\small +} \, {i \pi Emn \rho^2\over R} \bigr)$}}
\ee%\\[3.5mm]
%& \simeq & \Bigl({{\mbox{\raisebox{-1pt}{$\small T \rho$}}}\over
%{\mbox{\raisebox{-1pt}{\small $\pi  R$}}}}\Bigr) \; \sum_{m,\ell}\;
%e^{- {\textstyle {T\over 2}}
%\bigl( \bigl({\textstyle{m \over R}-E\ell)^2 + \ell^2/\rho^2\bigr)}} \nonumber
As the final step, we may now evaluate the integral over the
Schwinger parameter $T$ via the saddle point approximation. The
saddle points are at\footnote{We drop the term with $n=0$, since
it corresponds to the vacuum contribution.} \be T_0 = {2\pi \rho
\, |\, n|\over \sqrt{{m^2\over R^2} + \Lambda _\EE k_i^2}}~,
\hskip 1cm n=\, \pm 1 \, , \, \pm 2 \, , \, \ldots ~. \ee Before
plugging this back in to obtain our final result, let us first
briefly check our assumption that $\rho^2 \! >\!\!>\! T$: setting
$k_i=0$, we find $\rho^2/T _0= m\rho/2\pi |n| R$. So as long as
the spatial distance scale $\rho$ is much larger than the KK
compactification radius $R$, we're safe to use (\ref{villain}).

With this reassurance, we proceed and find our final answer for
the pair creation rate per unit time and volume where we made the
replacement $E\to iE$.\footnote{Note that the same result can be
obtained by replacing the sum in (\ref{form}) by
(\ref{poisresum}), integrating over $T$ exactly, and then using
eqn (\ref{exup}) to approximate the resulting Bessel function.}
\be %\label{almostfamous}
\label{famous} {\cal W} (\rho) \, = \,
{1\over 2\pi ^2 \rho ^{3/2}}\, \int \! {
{\prod\limits^{{}_{{}^{\! d-2}}}  dk_i \over {(2\pi)^{d-2}}}} \,
\sum_{m} \, \left ({m^2\over R^2}+\Lambda k_i^2 \right )^{1/4}
\sum _{n=1}^\infty \, {1\over n^{3/ 2}}
 \,  \exp\Bigl[- 2\pi n\, \omeg(\rho, q, k_i)
\Bigr] \ee where $\omeg(\rho, q, k_i)$ is the potential energy
introduced in equation (\ref{rpot}). The summation over $n$ in
(\ref{famous}) can be seen to correspond to the ``winding number''
of the Euclidean trajectory around the periodic Euclidean time
direction. The $n=1$ term dominates, and is the result announced
in the Introduction. Before discussing it further, we will now
proceed with a second method of derivation.

% = {\rho\over \Lambda}
% {\sqrt{{\textstyle{q^2\over R^2} + \Lambda k_i^2}}}\; + \;
% {q \, E \rho^2  \over 2 \Lambda R}
%= {|q|\, \rho \over 1\, -\,   {q\over |q|}\, E \rho}

\subsection{The Hawking-Unruh effect}\label{hawkingunruh}
The result (\ref{famous}) looks like a thermal partition function,
indicating that it can be understood as produced via the Hawking-Unruh
effect. We will now make this relation more explicit.

The functional integral (\ref{psum}) over all Euclidean paths
represent the one-loop partition function of a scalar field $\Phi$
in the d+1-dimensional electric KK Melvin space-time. We can
compute this determinant also directly via canonical quantization
of this field. The full expansion of $\Phi$ into modes starts with
a decomposition over wave numbers along the extra dimension  (in
this section we restrict $q$ to be positive) \be \Phi =\Phi_0 +
\sum\limits_{q = 1}^\infty  \Bigl(e^{i q x_{d+1}} \Phi _{q} +
e^{-iq x_{d+1}} \Phi^*_q\Bigr) \; , \ee where $\Phi_0$ is massless
and real, and $\Phi_q$ are complex and have mass $m=q$. Let us
define $\mu_+$ and $\mu _-$ via \be \mu_+^2 =  (q + \half \omega
E)^2+k_i^2 \ , \qquad \quad \mu _-^2=(-q+\half \omega E)^2+k_i^2
\; , \ee so that now $\mu _\pm$ are quantities related to
positively or negatively charged particles.

To proceed, we now need to expand the field $\Phi_q$ in creation and
annihilation modes, allowing only
modes that satisfy the boundary conditions (\ref{bc}) that $\Phi_q(\rho_i)=0$
at the location of the two charged plates.
\be
\label{expand}
\Phi _{q}=\, \sum\limits_{\omega>0}^{} \;
\int \!\!  {\prod\limits^{{}_{{}^{}_{\!d-2}}}  dk_i
\over (2\pi)^{{d-\!2}}} \,
{e^{-i\omega \tau+ i k_i y^i}\over \sqrt{\omega f(\omega)}} \;
\Big( \, K(\omega,\mu _+ \rrr)\,
 a_{q} (k_i, \omega) + K^*(\omega,\mu _- \rho)\,
a_{-q}^\dagger(k_i, \omega)
\Big)\  ,
\ee
with $K(\omega,\mu\rho)$ and $f(\omega)$ as defined in (\ref{kdef})
and (\ref{inn}). The creation and
annihilation modes then satisfy the usual commutation relations.
\be [a_{\pm q}(k_1,\omega_1),
a_{\pm q}^\dagger(k_2,\omega_2)]=\delta (k_1\!-\!k_2)
\delta_{\omega_1\omega_2}.
\ee

Our goal is to determine what the natural vacuum state of the $\Phi$ field
looks like, as determined by the initial conditions. In the far past,
we imagine that the KK electric field was completely turned off. The
electric KK Melvin background then reduced to Rindler or Minkowski
space -- depending on which coordinate system one introduces. The most
reasonable initial condition is that the quantum state of all $\Phi$ quanta
starts out in the vacuum as defined in the Minkowski coordinate system.
Let us denote this Minkowski vacuum by $|\Omega\rangle$.

To determine the expression for $|\Omega\rangle$ in terms of our
mode basis, we can follow the standard procedure \cite{Unruh}
\cite{BD}. We will not go into the details of this calculation
here, except to mention one key ingredient: the mode functions,
when extended over the full range of $\rho$ values, have a
branch-cut at the horizon at $\rho=0$, such that
\be
\label{simple} K(\www,-\mu \rho )\,=\, e^{\pm 2 \piome}
K^*(\www,\mu \rho)\,,
\ee
depending on whether the branch cut lies in the upper or
lower-half plane. This behavior of $K(\omega,\mu \rho)$ near
$\rho\! =\! 0$ is sufficient to deduce the form of the Bogolyubov
transformation relating the modes $a(\omega,k)$ to the Minkowksi
creation and annihilation modes. (see e.g. \cite{BD}). As a
result, one finds that the Minkowski vacuum, $|\Omega\rangle$,
behaves like a thermal density matrix for the observable creation
and annihilation modes in (\ref{expand}). In particular, the
number operator for each mode has the expectation value
\be \label{thermal}
\langle\,
\Omega \, |\, a^\dagger_{q}(k,\omega)\, a_{q}(k,\omega)\,
| \, \Omega\, \rangle
%~\langle\Omega  |\, a_{L,q}\, a^\dagger_{L,q}\, |\Omega\rangle
~=~ {1\over e^{2\piome}-1}, \ee
while the overlap of $|\Omega\rangle$ with the empty vacuum state,
defined via $a_q(k,\omega)|\, 0\,\rangle$ =0, becomes
\bea
\label{thermal2}
|\langle \, 0 \, |\, \Omega \, \rangle|^2 &=& \exp\left[
 \int \!\! {\prod\limits^{{}^{{}_{\!d-2}}}  dk_i
\over (2\pi)^{{d-\!2}}} \sum_{q, \omega}
\, \Bigl | \log
\Bigl(1 - e^{- {2\piome}}\Bigr) \Bigr | \right]\\
&=& \exp\left[ - \int \!\! {\prod\limits^{{}^{{}_{\!d-2}}}  dk_i
\over (2\pi)^{{d-\!2}}} \sum_{q, \omega ,n}\, {1\over n}\,
e^{-2n\, \piome} \right ] .\eea This expression represents the
probability that the state $\Omega$ does not contain any particles
-- and its dominant $n=1$ term looks indeed closely related to the
result (\ref{famous}) obtained in the previous subsection.

The difference between the two equations is that (\ref{famous}) is
defined at a particular location $\rho$, while (\ref{thermal2})
contains a summation over all frequencies. To make the relation
more explicit, imagine placing some measuring device at a location
$\rho$. As mentioned before, only modes with a sufficiently large
frequency will reach this location with any appreciable
probability, and the probability attains a maximum for frequencies
equal to the potential energy at $\rho$, since for those
frequencies, $\rho$ is the turning point. Via this observation, we
can view the position $\rho$ as a parametrization of the space of
frequencies, via the insertion of \be 1 = \int \! d\rho \,
{\mbox{\large $\delta$}}\Bigl(\omega - \omeg(\rho)\Bigr) \;
|\partial_\rho \omeg(\rho)| ~,\ee with $\omeg(\rho)$ as given in eqn
(\ref{rpot}), thus replacing the summation over $\omega$ in
(\ref{thermal2}) by an integral over $\rho$. The integrand at
given $\rho$ is then naturally interpreted as the production rate
(\ref{famous}) at the corresponding location. This procedure is a
good approximation provided the distance $d\!=\! \rho_2\!-\rho_1$
between the plates is large enough, so that many frequencies
contribute in the sum.

This same condition is also important for a second reason
\cite{PB}. Since we would like to imagine that the pair production
takes place at a constant rate per unit time, we would like to see
that the overlap (\ref{thermal2}) in fact decays exponentially
with time. This comes about as follows \cite{PB}. Suppose we
restrict the field modes to be supported over a finite time
interval $0\! <\! \tau \!<\! T$. This translates into a
discreteness of the frequencies. Ignoring at first the other
discreteness due to the reflecting boundary condition at the two
parallel plates, it is clear that the density of frequencies
allowed by the time restriction grows linearly with $T$. The sum over
the frequencies thus produces an overall factor of $T$. In this
way, we  recover the expected exponential decay of the overlap
(\ref{thermal2}).

This exponential behavior breaks down, however, as soon as the time interval
$T$ becomes of the same order as the distance $d$ between the plates, or
more precisely, when $1/T$ approaches the distance between the discrete
energy levels allowed by the reflecting boundary conditions at the plates.
At this time scale, the situation gradually enters into a steady state,
in which the pair creation rate gets balanced by an equally large
annihilation rate. The system then reaches a thermal equilibrium, specified
by the thermal expectation value (\ref{thermal}). The physical
temperature of the final state depends on the location $\rho$ via
\be
\beta = 2\pi \sqrt{g_{{}_{00}}} = {2\pi \rho\over \sqrt{\Lambda}}.
\ee
Note that this temperature diverges at $\rho=0$ and $\rho=|2/E|$;
neither location is within our physical region, however.

\subsection{Charge Current}

It is edifying to consider the vacuum expectation value of the
charge current, since this is a clear physical, observer-independent
quantity and a sensitive measure of the local profile of the pair
creation rate. For given $q$, the charge current is given by \be j_\mu = i q
\Bigl( \Phi^*_q (\rho)\partial_\mu \Phi_q(\rho) - \Phi_q
(\rho)\partial_\mu \Phi^*_q(\rho)\Bigr) .\ee Using the result
(\ref{thermal}) for the expectation value of the number operator,
one finds that the time component of the current, the charge
density, is non-zero and equal to\footnote{Instead of the expectation
value (\ref{pcurrent}), one could also consider the mixed in-out
expectation value $\langle 0 | J_\tau(\rho) | \Omega\rangle$,
which is related to the derivative of the in-out matrix element
$\langle 0|\Omega\rangle$ with respect to $E$. This relation was
in fact used by Schwinger in his original derivation of the pair
creation rate \cite{Schwinger}.} \be \label{pcurrent} \langle \,
\Omega \, |\, j_{\tau}(\rho)\, |\, \Omega \, \rangle = J_+(\rho) -
J_-(\rho) \ee with \be \label{ccurrent} J_\pm(\rho) = q
\sum_{\omega >0} \int \!\!  {\prod\limits^{{}_{{}^{\!d-2}}}  dk_i
\over (2\pi)^{{d-\!2}}}\; {|K(\omega,\mu_\pm \rho)|^2 \, \over
f(\omega)\, \bigl(e^{2\piome} -\, 1\bigr)\, } \ee the positive and
negative charge contributions, respectively. Given the thermal
nature of the state $|\Omega\rangle$, the physical origin of this
charge density is clear: the presence of the electric field
reduces the potential energy of one of the two charge sectors,
thereby reducing its Boltzmann suppression, relative to the
oppositely charged.

To obtain a rough estimate for the behavior of $J_\pm(\rho)$, it
is useful to divide the frequency sum into three regions: i)
$\omega$ comparable to the potential energy (\ref{rpot}), ii)
$\omega$ much larger, or iii) $\omega$ much smaller. By comparing
the respective suppression factors, we find that the leading
semi-classical contribution comes from regime i); this is also
reasonable from a physical perspective, since these are the
particles that spend most time near $\rho$. Regime ii) is strongly
Boltzmann suppressed and clearly negligible compared to
contribution i), while regime iii) is suppressed because the
corresponding mode functions $K(\omega,\mu\rho)$ are exponentially
small at the location $\rho$, via (\ref{exup}). The leading
contribution of region i) is of order $e^{-2\piome(\rho,q,k_i)}$,
in accordance with the result (\ref{famous}) for the pair creation
rate {\small ${\cal W}$}$(\rho)$.

Since the mode functions
$K(\omega, \mu\rho)$ are real (they are the sum of an incoming and
reflected wave), the current in the $\rho$ direction appears to vanish.
The result (\ref{ccurrent}) for the charge density indeed looks static.
This static answer, however, can not describe the time-dependent
pair creation process. Recalling our discussion above, however, we can recover
this time-dependence by restricting the sum over only those frequencies
necessary to cover the finite time interval $0\! <\! \tau \!<\!T$. This is
a $T$ dependent subset, thus leading to a $T$ dependent (initially linearly
growing) charge density.
However, when $1/T$ becomes much smaller than the step-size in the allowed
frequency spectrum, the steady state sets in and the
charge density indeed becomes a static thermal distribution given by
(\ref{pcurrent})--(\ref{ccurrent}).

\section{Discussion}\label{discussion}

In this paper we have tried to make a systematic study of the Schwinger
pair production of charged Kaluza-Klein particles. Due to their characteristic
property that their mass is of the same order as their charge $q$, the pair creation requires
such strong KK electric fields that gravitational backreaction can not be ignored.
We have included this backreaction by means of the electric KK Melvin solution,
and shown that, in spite of the fact that the electro-static potential can not
be made to exceed the rest-mass of the KK particles, pair production
takes place at a rate given by (\ref{uschwinger}).

What is the physical mechanism that is responsible for the pair creation?
Our final answer (\ref{uschwinger}) includes both the KK electric field
and a gravitational acceleration $a$. It is instructive to compare this
result with the known rate \cite{Spindel} for Schwinger pair production in an
accelerated frame, as quoted in eqn (\ref{rschwinger}) in the Appendix.
Since in our case $a$ is bounded below by $E/2$, we can only directly compare the
two answers in the limit of small electric field. In that limit, if we expand the log in eqn (\ref{rschwinger}) and take the dominant $n=1$ term there, both answers become 
\be
\label{compare}
{\mbox{\small ${\cal W}$}}(E,a) \; \simeq \; \sum_{q = \pm |q|}
\int {\prod\limits^{{}_{{}^{\! d-2}}}  dk_i
\over {(2\pi)^{d}}} \exp\Bigl(-2\pi
\Bigl({1\over a} \sqrt{q^2 + k_i^2} -
{qE\over 2 a^2} \Bigr)\Bigr).
\ee
In this regime, however, one can not honestly separate the Schwinger pair
creation effect
from the pair creation effect due to the acceleration. Electric charge is
being produced, but it is just a simple consequence of the fact that the
electrostatic potential reduces the Boltzmann factor for one type of charge, while
increasing it for the other. Rather than producing the charge ``on its own,''
the electric field just polarizes the thermal atmosphere produced by the
Unruh effect.

In fact, if we write the potential $\omeg(\rho,q,k_i)$ as
in (\ref{rpot}) instead of (\ref{rpot2}),
our final answer (\ref{uschwinger}) appears to be just a small modification of
(\ref{compare}) and the physics that leads to it indeed seems quite identical.
So depending on taste, one can either interpret our result (\ref{uschwinger})
as pair creation due to a combination of the Schwinger and Unruh effect,
or as the result of the Unruh effect only.
There is no definite way to decide
between the two, since the gravitational acceleration can not be
turned off independently. Either way, what is clear is that the mechanism for
pair creation cannot be given a tunneling interpretation.

\bigskip

\bigskip

\noindent
{\bf \large Acknowledgements}

\vskip .2cm

\no We would like to thank O. Ganor, N. Itzhaki, K.T. McDonald, and A.M. Polyakov for helpful discussions. T.F. would also like to thank the organizers of the Cargese Summer School 2002, where some of this work was carried out.

\no  The work of T.F. is supported in part by a National Science Foundation Graduate Research Fellowship, in part by a Soros Fellowship for New Americans, and in part by the National Science Foundation under Grant No. PHY-9802484. The work of H.V. is supported by the National Science
Foundation under Grant No. 98-02484.

\no Any opinions, findings, and conclusions or recommendations expressed in
this material are those of the authors and do not necessarily reflect
the views of the National Science Foundation or the Soros Foundation.

\bigskip

\bigskip

\appendix
\no \begin{Large} {\bf Appendix} \end{Large}

\section{Gedanken Apparatus}\label{experimentalapparatus}

For a good understanding of the situation we wish to study, it
will be useful to investigate how, via a concrete Gedanken
experiment, one may in fact attempt to create a large static
Kaluza-Klein electric field. Without taking into account
gravitational backreaction, we imagine taking two parallel plates
with opposite KK charge density per unit area $\sigma$ and
perpendicular distance $d$, thus creating an electric field $E =
4\pi \sigma$ in the region between the plates. It turns out
however, that when we include the gravitational backreaction of
both the plates and the electric field, there are some
restrictions on how symmetric, or static, we can choose our
experimental set-up.

Consider two charged, infinitesimally thin, parallel plates at positions
$\rho _1$ and $\rho _2$, separated by a distance
\be
\label{distance}
d = \rho_2 -\rho_1~.
\ee
The two plates divide space into three regions:
Region A left of the first plate, given by $\rho <\rho _1$,
region B in between the two plates, $\rho _1 <\rho<\rho _2$, and region C
right of the second plate $\rho >\rho _2$.

Let the mass densities of the plates be given by $\mu_1$ and $\mu_2$, so that
\be
T^{0}_{\, 0} = \, \mu_1 \, \delta(\rho\!-\!\rho_1)
\, + \, \mu_2 \, \delta(\rho\!-\! \rho _2)~.
\ee
In addition, the two plates have charge densities $\sigma_1$ and $\sigma_2$
\be
\sqrt{g_{00}} \, \; T^0_{d+1} = \, \sigma_1 \, \delta(\rho\!-\!\rho_1)
\, + \, \sigma_2 \, \delta(\rho\!-\! \rho _2)~.
\ee
We will assume that the charge densities are opposite, $\sigma _1 = -\sigma _2$, and tuned so that
there is a Kaluza-Klein electric field in region B between the plates,
but none in regions A or C outside the plates. The region between the
plates therefore takes the form of a static slice $\rho_1 < \rho < \rho_2$
of the electric KK Melvin space-time. The two regions outside the
plates, on the other hand, are just flat. More precisely,
since the parallel plates in effect produce an attractive gravitational
force on freely falling particles in the two outside regions, the
regions A and C should correspond to static sub-regions in Rindler space.

Both Rindler space and the electric KK Melvin solution differ from
Minkowski space only via the $g_{00}$ component.
Imposing continuity at $\rho=\rho_i$, this leads us to the following Ansatz
for the $g_{00}$ component of the metric in the three regions
\bea
\label{goo}
g^A_{00}\!(\, \rho)\is \bigl(1\!-\!a_1 ({\rho}\!-\!\rho_1)\bigr)^2  \, g_{00}^B\!(\,\rho_1)\, ,
% -\rho _1 \over a} \Big )^2 /\Lambda _1~,
\nonumber \\[3mm]
g_{00}^B\!(\, \rho) \is {\rho^2 \, \over 1 -
E^2 \rho^2/4}  ~,\nonumber \\[3mm]
g_{00}^C\!(\, \rho) \is \bigl(1\!+\! a_2 ({\rho\! -\! \rho _2})\bigr)^2 \;
g_{00}^B\!(\,\rho_2).
\eea
Here $a_1$ and $a_2$ are both positive, and represent the respective free fall
accelerations of freely moving particles just outside of the two plates. In other
words, via the equivalence principle, $a_1$ and $a_2$ are the accelerations
(to the left and right, respectively) of the two plates as viewed from the outside
Minkowski observers. The quantities $\rho_1$ and $\rho_2$ play a similar role,
and can be both positive and negative. A physical restriction, however,
is that the denominator in the expression (\ref{goo}) for $g_{00}^B$ remains
positive.

In addition there is a non-trivial electric potential $g_{0,d+1}^B$ in the region
between the plates, while $g_{0,d+1}^{A,C}$ are constants determined by continuity:
\bea
\label{potential}
g_{0,d+1}^A\!(\,\rho) \is g_{0,d+1}^B\!(\,\rho_1) ~,%{E\over {1-E^2/b^2}}  ~,
\nonumber \\[3mm]
g_{0,d+1}^B\!(\,\rho) \is {\rho^2 E/2 \over 1 -  E^2 \rho^2/4}
 ~,\\[3mm]
g_{0,d+1}^C\!(\,\rho) \is g_{0,d+1}^B\!(\,\rho_2) .\nonumber
\eea
%where
%\bea && \Lambda = 1-E^2\Big (1+{\rho -\rho _1 \over b} \Big )^2 ~,\nonumber \\
%&&\Lambda _1=\Lambda (\rho _1)=1-E^2~,\nonumber \\
%&&\Lambda _2=\Lambda (\rho _2)=1-E^2N^2 ~, \nonumber \\
%&&N_E^2=\Big (1+{\rho _2 - \rho _1 \over b} \Big )^2 ~/\Lambda _2=N^2/\Lambda _2~.
%\eea
%As long as $a,b,c$ are positive and $1-E^2N^2>0$, the expressions for $g_{00}^{A,B,C}$ are always positive,
The d+1-dimensional Einstein equations of motion result in the following jump
conditions for the
normal variations of $g_{00}$ and $g_{0,d+1}$ at the location of the
plates\footnote{Note that while the expressions $g_{0,d+1}$ are not gauge invariant,
the Gauss equation is, as long as $\lambda$ in $A_\mu dx^\mu\rightarrow
A_\mu dx^\mu +d\lambda$ is smooth across $\rho _1$ and $\rho_2$,
i.e. $(\pa _{\rho _+}-\pa _{\rho _-})\lambda ~|_{\rho =\rho _i}=0~. $}
\bea
\label{israel}
4\pi \mu_i &\! = &\!
g^{00}(\pa _{\rho _+}g_{00}-\pa _{\rho _-}g_{00})~|_{\rho=\rho_i}, \\[3mm]
4\pi \sigma_i &\! =\! & \!
(g^{00})^{1\over 2}
(\pa _{\rho _+}g_{0,d+1}-\pa _{\rho _-}g_{0,d+1})~|_{\rho =\rho_i}~.
\eea
The first of these equations is known as the Israel equation, while the
second is equivalent to Gauss' law in electro-magnetism. Inserting
our Ansatz, the Israel jump conditions become
\bea
\label{israel2}
\qquad  \quad
2\pi \mu_1 \is  a_1 \, + \, {1 \over \rho_1\, \Lambda _1} \nonumber \\[-2mm]
& & \qquad \qquad \qquad \qquad \qquad \ \ \Lambda_i = 1 -\, \quart
\, E^2 \rho_i^2 \\[-2mm]
\qquad 2\pi \mu_2 \is \, a_2 \, -\,  {1\over \rho_2\,
\Lambda _2} \nonumber
\eea
while Gauss' law takes to the form
\be
\label{gauss2}
 \qquad 4 \pi \sigma_1  =  {E/ \Lambda^{3/ 2} _1}\ ,\qquad \quad \
4 \pi \sigma_2 = - {E/ \Lambda^{3/ 2} _2}\, .
\ee
Equation (\ref{israel2}) relates the mass density of the two plates to
the jump in the surface acceleration when moving from one to the other side,
while (\ref{gauss2}) relates the charge density to the jump in the KK
electric field.

Let us briefly check these formulae by considering some special
cases. If $E=0$, then we can choose the symmetric situation $\mu_1\!=\!\mu_2$
and $a_1=a_2$. Via (\ref{israel2}) this implies that we should take
the limit $\rho_i \! \to \! \infty$ keeping the distance (\ref{distance})
fixed. 
The intermediate region $B$
then simply reduces to
flat Minkowski space. This is as expected, since the two plates lead
to an equal and opposite gravitational force, which exactly cancels in the
intermediate region. For non-zero $E$, yet small electro-static potential
$V_{12} = E d$ between the plates, we can choose parameters such that
$E \rho_i <\!\!< 1$ and $\rho_i >\!\!> d$. Eqn (\ref{gauss2}) then reduces
to the standard Gauss law of Maxwell theory.

\figuur{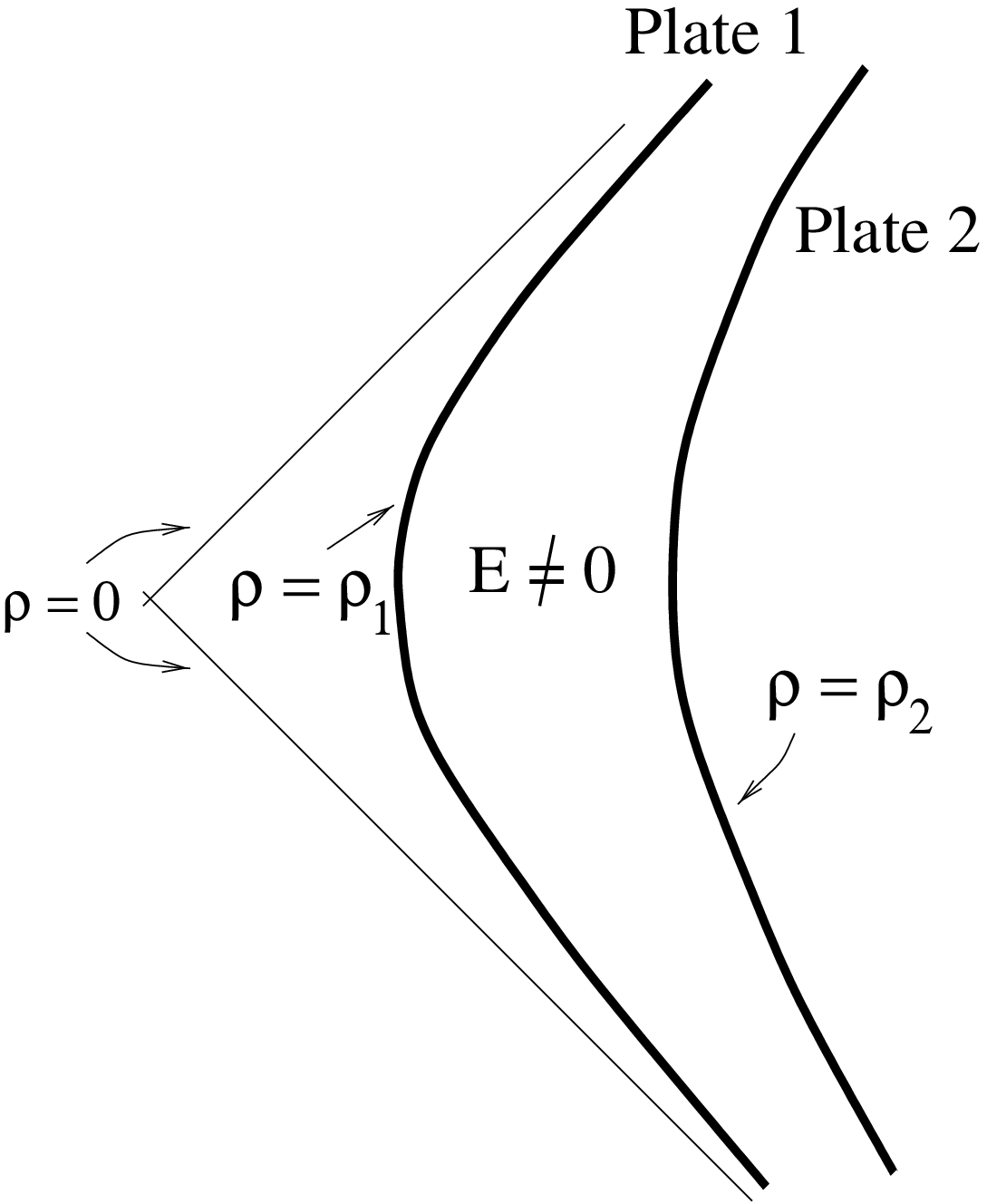}{6cm}{{\bf Fig  {\bf \ref{sit1}}: }{\ {\rm Situation I} \
describes the static situation
with two charged plates at $\rho \! = \! \rho_1$ and $\rho \! = \! \rho_2$,
with $0<\rho_1<\rho_2 < 2/E$.
The region of interest, in between the two plates,
is a static slice of region I of the electric KK Melvin solution.} }{sit1}

Let's now consider the general case.
There are four equations, and (for given inter-plate distance $d$,
and densities $\mu_i$ and $\sigma_i$ ) four unknowns: $a_1$, $a_2$, $\rho_1$
and $E$. The second  equation in (\ref{gauss2}), however, is not really
independent from the
first, since we should rather read it as a fine-tuning condition on
$\sigma_2$ (relative to $\sigma_1$) ensuring that the $E$-field
vanishes outside the two plates. Discarding this equation, we are
thus left with one overall freedom, namely, the overall acceleration
of the center of mass of our apparatus.

For practical purposes, we would have preferred to restrict
ourselves to the simplest and most symmetric case in which the two plates
have equal mass density $\mu_1\! = \! \mu_2$ and equal
surface acceleration $a_1 \! =\! a_2$. This would in particular ensure that
our apparatus is at rest. As seen from eqn (\ref{israel2}), this
symmetric situation could be reached if we could take
the limit $\rho_i \! \to\! \infty$.
However, for non-zero $E$, this limit is forbidden via the restriction
$1\!-\! \quart
E^2 \rho^2 >0$. Thus we are basically forced to consider the general
situation with $\mu_i$ and $a_i$ arbitrary, and $\rho_i$ both positive.
We call this:
\be
\label{sitI}
{\rm Situation\ I}: \qquad 0<\rho_1< \rho_2 < 2
/E, \qquad \quad \mu_i\ \ {\rm arbitrary}
\ee
In this case, the region of interest, region B, represents a static slice in
the right wedge of the electric KK Melvin solution. This Situation I is
the natural generalization of a constant, static electric field, and is our
starting point for studying the possible Schwinger pair creation
of charged KK particles. We sketch it in Fig. 5.

\figuur{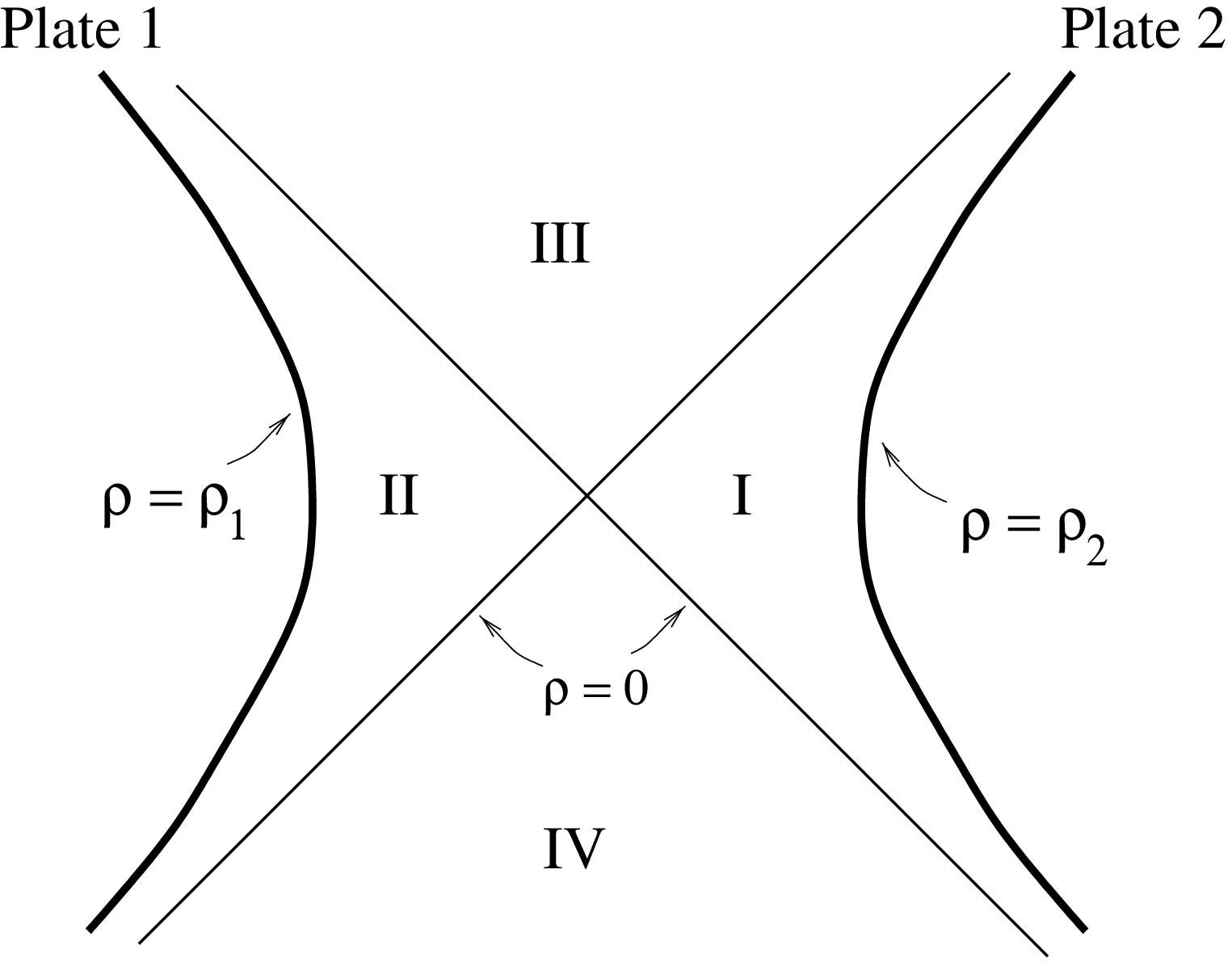}{6.6cm}{{\bf Fig {\bf \ref{sit2}}: }{Situation II describes the time-dependent
situation with two accelerating
charged plates at $\rho \! = \! \rho_1 < 0$ and
$\rho \! = \! \rho_2 > 0$. The region of interest, in between the two plates,
includes the time-dependent regions III and IV of the electric KK Melvin
solution.} }{sit2}

There is, however, another situation we could consider, which does
allow for a symmetric solution. Namely we can choose:
\be
\label{alternative}
{\rm Situation\ II}: \qquad \mu_1 =  \mu_2\, , \qquad \ \ a_1 = a_2 \, , \qquad \ \
\rho_1 = - \rho_2 .
\ee
In this case the region B includes the special position $\rho = 0$ at
which $g_{00}=0$, the location of the event horizon of the electric
KK Melvin geometry (see Fig. 6).

To better understand the experimental conditions leading to
Situation II, consider the special case $\mu_1\!=\!\mu_2\!=\!0$,
and $\sigma_1\! =\!\sigma_2\!=\! 0$. This describes two plates
with zero mass and charge, accelerating away from each other with
equal but opposite acceleration $a_i = 1/\rho_i$. It is now easy
to imagine that one can gradually add mass and charge to the
plates, and reach the general Situation II. It must be noted that
this experimental set-up does not lead to a static background,
since the geometry now includes the time-dependent regions III and
IV enclosed by the Rindler horizon (see fig \ref{sit0}). This
set-up is therefore not a direct analog of the static electric
field considered by Schwinger. For a discussion of Situation II
see \cite{CC}; our main focus is Situation I.

\section{Schwinger meets Rindler}\label{SKK-SR}

In this Appendix we summarize the known result for the Schwinger
pair creation rate in an accelerating frame \cite{Spindel} of
charged particles with mass $q$ and mass $m$ with $m\! <\!\!<\!
q$. In this regime, pair creation starts to occur while the
gravitational backreaction of the electric field is still
negligible.

The closest analog of a constant, uniform gravitation field is
Rindler space
\be
\label{un}
 ds^2 = - \rho^2 d\tau^2 + d\rho^2 + dy_i^2.
\ee
Particles, or detectors, located at a given $\rho$ undergo a
uniform acceleration $a=1/\rho$. Consider a charged field
propagating in this space in the presence of a uniform electric
field, described by
\be
\label{deux}
 A_\tau = {1\over 2} E \rho^2.
\ee
The resulting scalar wave equation reads
\be
\label{trois}
\label{gcen}\left( \frac{1}{\rho}\,
\partial_\rho \bigl({\rho} \,\partial_\rho \bigr)\, -\, \frac{1}{\rho ^2}\Bigl(
\pa _{\ttau} + {i\over 2} q E\rho^2\Bigr)^2 \, \, + \, \pa_i^2 \,
- \, m^2 \right)\Phi= 0\, . \ee The above three equations are
connected to the ones in Section \ref{waveeqns} by setting
$\Lambda\!=\!1$ (which indeed amounts to turning off the
backreaction) and by setting $m\!=\! q$ above.

Eqn (\ref{trois}) has known mode solutions with given Rindler
frequency, in terms of Whittaker functions \cite{Spindel, GR}. These functions have a relatively intricate, but known
(eqn 9.233 in \cite{GR}), branch-cut structure at $\rho\!=\!0$,
from which one can straightforwardly extract the linear
combination of (left and right wedge) Rindler creation and
annihilation modes that annihilate the Minkowski vacuum
$|\Omega\rangle$. One obtains the following result for the total
pair creation rate per unit time and (transverse) volume
\cite{Spindel}
\be
\label{rschwinger}
{\mbox{\footnotesize ${\cal W}$}} \; \simeq \; \,
\sum_{q = \pm |q|}
 \int \!\! d\omega \, \int {\prod\limits^{{}_{{}^{\! d-2}}}  dk_i
\over {(2\pi)^{d}}}
\; \log\,\left(
{\Bigl( \, 1 -  e^{-2{\mbox{\small$\pi$}}
{\mbox{\small $\omega$}}}\Bigr)\Bigl(1-  e^{-
\textstyle  {\pi (m^2 + k_i^2) \over |qE|}}\Bigr)
\, \over 1-  e^{-2{\mbox{\small$\pi$}}\bigl(\ome
+ {\textstyle  {m^2 + k_i^2 \over 2|qE|}}\bigr)}}\right).
\ee
As explained in Section \ref{hawkingunruh}, we can extract from this result the
pair creation rate at a given radial location $\rho$, or
equivalently, given acceleration $a\!=\! 1/\rho$, by equating the
frequency $\omega$ with the classical potential energy at this
location 
\be
\label{ccpot}
\omega(q,k_i) = {1\over a} \sqrt{q^2 + k_i^2} -
{qE\over 2 a^2}.
\ee
As discussed in section \ref{discussion}, in the limit where the electric field is small, the expression (\ref{rschwinger}) reduces to our result.

\end{document}